\documentclass[a4paper,12pt,oneside]{article}
\usepackage[latin1]{inputenc}
\usepackage{amsmath}
\usepackage{amsfonts}
\usepackage{amssymb}
\usepackage{graphicx}
\usepackage{footnote}
\usepackage{float}
\usepackage{subcaption}
\DeclareGraphicsExtensions{.bmp,.png,.pdf,.jpg,.eps}
\usepackage{physics}
\usepackage{mathrsfs}

\usepackage[colorlinks]{hyperref}
\hypersetup{
    colorlinks=true,
    linkcolor=blue,
    filecolor=magenta,  
    urlcolor=[rgb]{0.00,0.00,0.50},    
    citecolor=red,
    }

\usepackage{url}

\advance \textheight by \topskip
\usepackage{amsmath}
\usepackage[english]{babel}
\makeatletter
 \@addtoreset{equation}{section}
\makeatother
\usepackage[latin1]{inputenc}
\usepackage{amsmath}
\numberwithin{equation}{section}

\advance \textheight by \topskip
\usepackage{amsmath}
\usepackage[english]{babel}
\DeclareMathAlphabet\mathbfcal{OMS}{cmsy}{b}{n}
\DeclareMathAlphabet{\boldmathe}{T1}{cmr}{bx}{it}
\newcommand{\mbf}[1]{\boldmathe{#1}}

\def\vx{\mbf{x}}

\def\be{\begin{equation}}
\def\ee{\end{equation}}

\def\R{\mathbb R}

\def\C{\mathbb C}

\def\be{\begin{equation}}
\def\ee{\end{equation}}

\def\R{\mathbb R}

\def\C{\mathbb C}

\def\be{\begin{equation}}
\def\ee{\end{equation}}

\def\Z{\mathbb Z}

\def\R{\mathbb R}

\def\C{\mathbb C}

\usepackage{todonotes}

\begin{document}

\title{
{\bf{
Conformal generation of an exotic  rotationally invariant
harmonic
oscillator }}
}

\author{{\bf  Luis Inzunza and Mikhail S. Plyushchay} 
 \\
[8pt]
{\small \textit{Departamento de F\'{\i}sica,
Universidad de Santiago de Chile, Casilla 307, Santiago,
Chile  }}\\
[4pt]
 \sl{\small{E-mails:   
\textcolor{blue}{luis.inzunza@usach.cl},
\textcolor{blue}{mikhail.plyushchay@usach.cl}
}}
}
\date{}
\maketitle

\begin{abstract}
An exotic  rotationally invariant  harmonic oscillator (ERIHO)
 is constructed by applying a non-unitary isotropic 
conformal bridge
 transformation (CBT)
 to a free planar  particle. 
It is described by the  
 isotropic harmonic oscillator Hamiltonian  supplemented  by  a Zeeman type term 
with a real coupling constant $g$.
The model reveals  the  
Euclidean  ($|g|<1$) 
and Minkowskian  ($|g|>1$)  phases 
separated by the phases $g=+1$ and $g=-1$  of 
 the Landau problem in the symmetric gauge
with opposite orientation of the magnetic field.
A hidden symmetry emerges in the system 
at rational values of  $g$.
 Its generators, together with the Hamiltonian 
and angular momentum produce non-linearly deformed
$\mathfrak{u}(2)$ and $\mathfrak{gl}(2,\R)$ algebras
in the cases of $0<|g|<1$ and $\infty>|g|>1$, 
which transmute one into another
under the inversion
$g\rightarrow -1/g$.
Similarly, the true, $\mathfrak{u}(2)$, 
and 
extended conformal, 
$\mathfrak{gl}(2,\R)$,
symmetries  of the
isotropic Euclidean oscillator ($g=0$) interchange their 
roles  in  the isotropic Minkowskian oscillator ($|g|=\infty$), 
 while  two copies of   the $\mathfrak{gl}(2,\R)$ algebra 
 of  analogous symmetries mutually  transmute in Landau phases.
  We show that the ERIHO system  is transformed by a  
  peculiar unitary transformation into the  anisotropic harmonic oscillator
  generated, in turn,  
  by anisotropic CBT.
  The relationship between the ERIHO and the subcritical phases of the 
  harmonically extended Landau problem, as well as with a plane isotropic harmonic 
  oscillator in a uniformly rotating reference frame, is established.
\end{abstract}

\section{Introduction}

The revival of interest in non-relativistic conformal symmetry \cite{AFF,SCM5}
was stimulated by non-relativistic AdS/CFT correspondence 
\cite{LeiPly,BarFue,DuvHasHor,Jack},
its relevance to black holes physics and cosmology 
\cite{ConformalBH1,ConformalBH2,ConformalBH3,DGH,ConformalBH0, PioWal,CGGH},
and 
its  utility in  the description of 
 strongly coupled condensed matter systems \cite{GAdS1,GAdS2,App2,App3} and  
  QCD confinement problem \cite{App1,Brod2}.
 In this context, the mechanism to improve 
 the properties of the scale-free conformal mechanics proposed 
initially by de Alfaro, Fubini and Furlan  \cite{AFF} amounts to 
an improved choice of the  time coordinate in black holes physics
since  a usual time variable is not a good global evolution coordinate 
on AdS${}_2$ \cite{ConformalBH0}.  Via the same  basic mechanism, 
the mass and length scales are introduced in holographic QCD \cite{Brod2}.
On the other hand, this mechanism 
corresponds to the Niederer's transformation \cite{NH1},
by which the relation between the free particle's 
and harmonic oscillator's dynamics 
was established at the classical and quantum levels.
The latter relationship, it turn, 
corresponds to different forms of dynamics  \cite{Dirac} 
with respect to the conformal symmetry. 

In recent papers \cite{InzPlyWipf1,InzPlyWipf2,InzPly7},
the non-unitary conformal bridge transformation 
(CBT)
was introduced,  by which  the non-compact and compact 
generators of the conformal symmetry can be related
 in the spirit of Dirac's different forms of dynamics.
This allowed to establish the relation 
between  the quantum states and symmetries,
including hidden symmetries,  of different 
 asymptotically free and 
associated harmonically trapped systems.
The correspondence comprises not only  energy 
eigenstates of the systems, but  also 
coherent and squeezed states. 
The non-unitary CBT turns out to be  closely 
related with a unitary transformation between 
the quantum coordinate (Schr\"odinger) and holomorphic
(Fock-Bargmann) representations. Its classical analog 
yields  a canonical transformation  
corresponding to the Hamiltonian vector flow produced by
generators of the conformal symmetry taken 
with particular complex values of the parameters.
Bearing in mind these two last properties, 
the CBT shows some not explored yet similarity 
with the $\mathcal{PT}$-symmetry \cite{Bender,Mostafazadeh,Bender2007,Fring,DCT}. 

The  CBT employed in refs. \cite{InzPlyWipf1,InzPlyWipf2,InzPly7}
 in different geometric and dynamical
backgrounds possesses the property 
of rotational invariance. In this work, we exploit 
the isotropy of the CBT to generate 
the exotic  rotational invariant 
 one-parametric  family 
of  harmonic oscillator  systems
which have a number of properties
 interesting from a physical point of view.
As will be showed, in dependence on the 
value of the real parameter $g$, 
 the family reveals two distinct phases 
with Euclidean ($|g|<1$) and Minkowskian 
($|g|>1$) properties, which  are separated 
by the phases $g=+1$ and $g=-1$ corresponding to
 the Landau problem in the symmetric gauge
with opposite  orientation of the magnetic field.
At rational values of $g$, each  system of the family
possesses a hidden symmetry  \cite{Cariglia}, 
and so, is  maximally super-integrable.
Symmetries of  different phases mutually transmute  
 under  the inversion
 $g\rightarrow -1/g$, and  in general case they
 generate  non-linear algebras of the W-type \cite{BHT}.
 We also reveal a  unitary relationship  of the exotic rotationally invariant harmonic oscillator 
 (ERIHO) 
 with the anisotropic harmonic oscillator (AHO) systems,
 and establish the relation  of our model with subcritical 
 phases of the harmonically extended Landau problem.

The paper is organized as follows.  In Sec. \ref{SecCBT}, 
the basic aspects associated with the 
CBT  are reviewed. 
In Sec. \ref{SecRIAHO}, 
 the non-unitary isotropic  CBT  is applied to  
 a complex linear combination of the generators of dilatations and rotations
of the two-dimensional free particle to produce 
the ERIHO.
The dynamics and symmetries of the system 
are discussed in detail  there at the classical and quantum levels.
 In Sec. \ref{SecAHOINCBT}, 
the anisotropic CBT  is used to generate 
the AHO
from the free particle.
In Sec. \ref{SecOtherS}, the relationships  of  the ERIHO with the AHO,
harmonically extended  Landau problem, and 
a plane isotropic harmonic oscillator in a uniformly 
 rotating reference frame are explored.
The final Sec. \ref{SecDisOut} is devoted to the discussion and outlook.

\section{The conformal bridge transformation}
\label{SecCBT}
The technique of  CBT   
\cite{InzPlyWipf1}
allows us to relate asymptomatically free systems (such as a free particle, or  conformal mechanics model),
characterized by the
$\mathfrak{so}(2,1)$  conformal symmetry,
with harmonically confined models that are $ \mathfrak{sl}(2,\R) $ conformal invariant 
(such as the harmonic oscillator, or the conformal mechanics model of de Alfaro, Fubini and  Furlan \cite{AFF}). 
In this section we review
the basic aspects associated with this transformation
to apply it then for the construction of the ERIHO.

\vskip0.1cm
Consider the quantum conformal $\mathfrak{so}(2,1)$
 algebra 
 \be
[\hat{D},\hat{H}]=i\hbar\hat{H}\,,\qquad
[\hat{D},\hat{K}]=-i\hbar\hat{K}\,,\qquad
[\hat{K},\hat{H}]=2i\hbar \hat{D}\,, \label{so(2,1)}
\ee
where $\hat{H}$, $\hat{D}$ and $\hat{K}$ are implied to be, respectively,  
the Hamiltonian, 
the dilatation generator, and 
the generator of special conformal transformations of a 
system, and we 
assume  here
that the operators 
$\hat{D}$ and $\hat{K}$  do not 
 depend explicitly on time.
By taking 
 linear 
combinations 
\be\label{JJJ}
\hat{\mathcal{J}}_0=\frac{1}{2\omega\hbar}(\hat{H}+\omega^2\hat{K})\,,\qquad
\hat{\mathcal{J}}_\pm=-\frac{1}{2\omega\hbar}(\hat{H}-\omega^2\hat{K}\pm 2i\omega\hat{D})\,,\qquad
\ee
we produce the $\mathfrak{sl}(2,\R)$ algebra 
\be
[\hat{\mathcal{J}}_0,\hat{\mathcal{J}}_\pm]=\pm \hat{\mathcal{J}}_\pm\,,\qquad
[\hat{\mathcal{J}}_-,\hat{\mathcal{J}}_+]=2\hat{\mathcal{J}}_0\,.
\ee
Though both algebraic structures are isomorphic, 
from a physical point of view 
they 
describe the systems with essentially 
different
properties
if $\hat{\mathcal{J}}_0$ is identified as 
the
 Hamiltonian of another model. The 
generator $ \hat{H} $ of  the $\mathfrak{so} (2,1) $ 
symmetry 
is non-compact and  has
a continuous spectrum. It  represents the Hamiltonian of an 
asymptomatically  free (for $|{x}|\rightarrow \infty$) particle.
 On the other hand, the operator   
$ \hat{\mathcal{J}}_0 $ 
is a compact generator of  the $ \mathfrak{sl}(2,\R)$ 
Newton-Hooke symmetry \cite{NH1,NH2,NH3,Papan,NH4}.
It is
characterized by a discrete spectrum 
corresponding  to  
 the associated
harmonically trapped system. The Planck 
constant $\hbar$ and parameter $\omega>0$ of the dimension of frequency introduced in  (\ref{JJJ}) guarantee
 the
 dimensionless character of  
 the generators $ \hat{\mathcal{J}}_0 $ and $ \hat{\mathcal{J}}_\pm $.

The \emph{non-unitary}  
operators 
\begin{eqnarray}
&
\hat{\mathfrak{S}}=e^{-\frac{\omega}{\hbar} \hat{K}} 
e^{\frac{\hat{H}}{2\hbar \omega}}
e^{\frac{i}{\hbar}\ln(2)\hat{D}}=
e^{-\frac{\omega}{\hbar} \hat{K}}e^{\frac{i}{\hbar}\ln(2)\hat{D}} e^{\frac{\hat{H}}{\hbar \omega}}\,,\qquad
\hat{\mathfrak{S}}^{-1}=e^{-\frac{i}{\hbar}\ln(2)\hat{D}}e^{-\frac{\hat{H}}{2\hbar \omega}}
e^{\frac{\omega}{\hbar}\hat{K}}\,,
&\label{QCB}
\end{eqnarray}
relate  (intertwine)  the sets of generators 
($\hat{H}$, $\hat{D}$, $\hat{K}$) 
and ($\hat{\mathcal{J}}_-$, $\hat{\mathcal{J}}_0$,
$\hat{\mathcal{J}}_+$)
by a 
similarity transformation, 
\begin{eqnarray}
&\label{GenConBr}
\hat{\mathfrak{S}}(\hat{H})\hat{\mathfrak{S}}^{-1}=-\omega\hbar \hat{\mathcal{J}}_-\,,\qquad
\hat{\mathfrak{S}}(i\hat{D})\hat{\mathfrak{S}}^{-1}=\hbar \hat{\mathcal{J}}_0\,,\qquad
\hat{\mathfrak{S}}(\hat{K})\hat{\mathfrak{S}}^{-1}=\frac{\hbar}{\omega}
\hat{\mathcal{J}}_+\,.&
\end{eqnarray}
The systems described by 
the
Hamiltonians $\hat{H}$ and $\hat{\mathcal{J}}_0$ 
 correspond, according to Dirac \cite{Dirac}, 
  to two different forms of dynamics associated here with 
 conformal symmetry. This is the quantum version of the 
CBT,
for some earlier applications of which see 
Refs. \cite{InzPlyWipf1,InzPlyWipf2,InzPly7}. 
This transformation implies,  in particular,  that 
\begin{eqnarray}
&\label{EigenEstates}
\hat{D}\ket{\lambda}=i\hbar \lambda \ket{\lambda}\quad \Rightarrow\quad \hat{\mathcal{J}}_0
(\hat{\mathfrak{S}}\ket{\lambda})=\lambda\hat{\mathfrak{S}}\ket{\lambda}\,,
&\\&
\label{CoherentEstates}
\hat{H}\ket{E}=E\ket{E}\quad \Rightarrow\quad \hat{\mathcal{J}}_-(\hat{\mathfrak{S}}\ket{E})=-\frac{E}{\hbar 
\omega}\hat{\mathfrak{S}}\ket{E}\,.&
\end{eqnarray}
One sees  \cite{InzPlyWipf1,InzPly7} then  that 
to get normalizable eigenfunctions of
 the operator  $\hat{\mathcal{J}}_0$, the formal eigenvector 
$\ket{\lambda}$ of the operator $\hat{D}$ has to 
satisfy the following properties:
\begin{itemize}
\item[I.] The series  $\exp(\frac{\hat{H}}{2\hbar \omega})\ket{\lambda}=
\sum_{n=0}^{\infty}\frac{1}{n!(2\hbar \omega)^{n}} (\hat{H})^n\ket{\lambda}
$
 has  to  reduce to  a finite number of terms, i.e., $\ket{\lambda}$ should 
be a Jordan 
state
 of the operator $\hat{H}$ corresponding to zero energy \footnote{
The wave functions of generalized Jordan states 
corresponding to energy $\lambda$ 
satisfy relations of the form $P(\hat{H})\Omega_{\lambda}=\psi_{\lambda}$, 
where $\hat{H}\psi_{\lambda}=\lambda\psi_{\lambda}$ and $P(\eta)$ 
 is a polynomial 
\cite{CJP,CarPlyJ,InzPly3}.
Here
we consider  the Jordan states satisfying  
the relations $(\hat{H})^\ell\Omega_{\lambda}=\lambda\psi_{\lambda}$ with $\lambda=0$ for a certain 
natural number $\ell$.}. 
\item[II.] The  wave functions $\bra{\vx}\ket{\lambda}$
 must not have poles and 
have to be single-valued.
\end{itemize}
On the other hand, the
 eigenvectors $\ket{E}$ 
(physical, or  non-physical,
with complex eigenvalues in general case)
 of $\hat{H}$  are transformed into eigenvectors of 
the lowering operator 
$\hat{\mathcal{J}}_-$ of  the $ \mathfrak{sl}(2,\R)$  algebra. 
Therefore, the resulting 
eigenstates in (\ref{CoherentEstates}) 
are the coherent states of the system
with the Hamiltonian  $\hat{\mathcal{J}}_0$.

The classical analog of  the CBT 
is given by the \emph{complex} canonical transformation \cite{InzPlyWipf1,InzPly7}
\be
\label{ClasicalBrige}
\mathscr{T}(\tau,\beta,\delta,\gamma,t)= T_{2\omega \mathcal{J}_0}(\tau)\circ T_{\beta\delta\gamma} \circ T_H(-t)\,,
\ee
where 
\begin{eqnarray}&\label{TransCan}
\exp(\gamma F)\star f(q,p):=f(q,p)+\sum_{n=1}^\infty 
\frac{\gamma^n}{n!}\{F,\{\ldots,\{F,f\underbrace{\}\ldots\}\}}_{n}=:T_F(\gamma)(f)\,&
\end{eqnarray}
is a Hamiltonian flux generated by a phase space function $F$,  and 
\begin{eqnarray}
&\label{Tabg0}
T_{\beta\delta\gamma}:=T_{K_0}(\beta)\circ T_{H}(\delta) \circ T_{D_0}(\gamma)=
 T_{K_0}(\delta)\circ T_{D_0}(\gamma) \circ T_{H}(2\delta)\,,
&\\& \text{with}\quad 
\delta=\frac{i}{2\omega}\,,\qquad
 \beta=-i\omega \,,\qquad \gamma=-\ln 2\,.
&\label{betadeltagamma}
\end{eqnarray}
Here, $D_0=D|_{t=0}$ and $K_0=K|_{t=0}$,
and we assume that the generators of dilatations, $D$,  
and special conformal transformations, $K$, 
are explicitly  depending on time,\emph{ dynamical} integrals of motion 
satisfying a  
relation of the form 
 $\dot{A}=\{A,H\}+\frac{\partial A}{\partial t}=0$.  In correspondence 
 with this, 
in the composed Hamiltonian flux
(\ref{ClasicalBrige}), the first transformation $T_{H}(-t)$ removes
 the $t$ dependence in the dynamical integrals $D$ and $K$.
The second transformation
relates these  generators   at $t=0$ with the generators of the $\mathfrak{sl}(2,\R)$ algebra $\mathcal{J}_0$ 
and $\mathcal{J}_\pm$ taken at $\tau=0$ 
(this is the classical analog of the quantum similarity transformation presented above).  
Finally,  $T_{2\omega \mathcal{J}_0}(\tau)$
restores the $\tau$ dependence {}\footnote{
We are interested in time dependence  of the classical dynamical integrals as
we will investigate the classical dynamics of the  corresponding systems.
The time dependence  in the  operators in Heisenberg picture can be restored analogously by inclusion of the respective 
evolution operators, but we will not be interested in it
at the quantum level.
} 
 of the generators $\mathcal{J}_\pm$. 
\vskip0.1cm
 
In particular case 
  of the $d$-dimensional quantum  free particle, its
 conformal symmetry generators 
 are given 
in the Schr\"odinger 
representation by
\begin{eqnarray}
&
\hat{H}=\sum_{i=1}^{d}\hat{H}_i\,,\qquad
\hat{D}=\sum_{i=1}^{d}\hat{D}_i\,,\qquad
\hat{K}=\sum_{i=1}^{d}\hat{K}_i\,,\label{TotalSo(2,1)}
&\\&
\hat{H}_i=\frac{-\hbar^2}{2m}\frac{\partial^2}{\partial x_i^{2}}\,,\qquad
\hat{D}_i=-i\frac{\hbar}{2}\left(x_i\frac{\partial}{\partial x_i}+\frac{1}{2}\right)\,,\qquad
\hat{K}_i=\frac{m}{2}x_i^2\,.\label{PartialSo(2,1)}
\end{eqnarray}
Here,  each set of generators $\hat{H}_i$, $\hat{D}_i$ and $\hat{K}_i$ 
satisfies the 
$\mathfrak{so}(2,1)$ algebraic  relations
(\ref{PartialSo(2,1)}), and so, the 
operators   $\hat{H}$, $\hat{D}$ and $\hat{K}$
 generate the same $\mathfrak{so}(2,1)$ Lie algebra. 
  
By using 
generators (\ref{TotalSo(2,1)})
to construct the conformal bridge operators (\ref{QCB}), one gets 
\be
\label{MultCBT}
\hat{\mathfrak{S}}=\Pi_{i=1}^{d}\hat{\mathfrak{S}}_i\,,\qquad
\hat{\mathfrak{S}}_i=e^{-\frac{\omega}{\hbar} \hat{K}_i} 
e^{\frac{\hat{H}_i}{2\hbar \omega}}
e^{\frac{i}{\hbar}\ln(2)\hat{D}_i}\,,\qquad
[\hat{\mathfrak{S}}_i,\hat{\mathfrak{S}}_j]=0\,.
\ee
 The CBT  produced by the operator $ \hat{\mathfrak{S}} $
  and its inverse is a composition of $d$ independent transformations. 
Each of these transformations touches a particular 
spatial direction, leaving the rest invariant, and 
here we use
the same parameter $\omega$ to guarantee the rotational invariance of the  transformation.
Later we will consider
other possibilities.
In correspondence with (\ref{GenConBr}), the total transformation 
$\hat{\mathfrak{S}}$
produces the generators  
\begin{eqnarray}
&
\hat{\mathcal{J}}_0=\sum_{i=0}^{d}\hat{\mathcal{J}}_0^{i}=\frac{1}{2\omega\hbar}\hat{H}_{\text{osc}}\,,\qquad
\hat{\mathcal{J}}_\pm=\sum_{i=0}^{d}\hat{\mathcal{J}}_\pm^{i}\,,&\\&
\hat{\mathcal{J}}_0^{i}=
\frac{1}{2} (\hat{a}_{i}^{+}\hat{a}_{i}^{-}+\frac{1}{2})=\frac{1}{2\omega\hbar}\hat{H}_{\text{osc}}^{i}\,,\qquad
\hat{\mathcal{J}}_\pm^{i}=\frac{1}{2}(\hat{a}_i^{\pm})^{2}\,,
&
\end{eqnarray}
where $\hat{H}_{\text{{osc}}}=\sum_{i=1}^{d} \hat{H}_{\text{{osc}}}^{i}$ 
is the quantum Hamiltonian of the $d$-dimensional isotropic harmonic oscillator, 
and 
\begin{eqnarray}
&\label{d-lader1}
\hat{a}_i^\pm=\sqrt{\frac{m\omega}{2\hbar}}\left(x_i \mp \frac{\hbar}{m\omega}\frac{\partial}{\partial x_i}\right)\,,\qquad
[\hat{a}_i^\pm,\hat{a}_j^\pm]=0\,,\qquad [\hat{a}_i^-,\hat{a}_j^+]=\delta_{ij}\,, 
&
\end{eqnarray}
are the usual first-order ladder operators of the system. 

In the free particle system, we also have the linear momenta operators and the Galilean boosts 
for each direction,
\be
\hat{p}_j=-i\hbar\frac{\partial}{\partial x_j}\,,\qquad
\hat{\xi}_{j}= m x_j\,,\qquad
[\hat{\xi}_j,\hat{\xi}_k]=[\hat{p}_j,\hat{p}_k]=0\,,\qquad
[\hat{\xi}_j,\hat{p}_k]=i\hbar m \delta_{jk}\,.
\ee
 The application 
of the CBT  to these operators produces (no summation  over
repeated index) 
\begin{eqnarray}
&
\hat{\mathfrak{S}}(\hat{p}_j)\hat{\mathfrak{S}}^{-1}=\hat{\mathfrak{S}}_j(\hat{p}_j)\hat{\mathfrak{S}}_j^{-1}=
-i\sqrt{m\hbar\omega}\hat{a}_j^-\,,&\\&
\hat{\mathfrak{S}}(\hat{\xi}_{j})\hat{\mathfrak{S}}^{-1}=\hat{\mathfrak{S}}_j
(\hat{\xi}_{j})\hat{\mathfrak{S}}_j^{-1}=\sqrt{\frac{m\hbar}{\omega}}\hat{a}_j^+\,.
&
\end{eqnarray}
The angular momentum tensor 
\be
\hat{M}_{ij}=\frac{1}{m}(\hat{\xi}_i\hat{p}_j-\hat{\xi}_j\hat{p}_i)=-i\hbar(\hat{a}_i^{+}\hat{a}_j^{-}-\hat{a}_j^{+}\hat{a}_i^{-})
\ee
commutes with the operator 
$\hat{\mathfrak{S}}$ 
and its inverse. 

As any symmetry generator of the $d$-dimensional 
free particle (harmonic oscillator) is a function 
of $\hat{\xi}_i$ and $\hat{p}_i$ ($\hat{a}_i^\pm$),
the CBT allows us to map the integrals of
 one 
 system to those 
of another system.

%
\section{ Exotic rotationally invariant harmonic oscillator}
\label{SecRIAHO}
%

From now on
we restrict ourselves to the case 
of
$d = 2$ 
and  
extend the CBT
of the previous section to generate and investigate the 
ERIHO. 
For this we exploit the invariance of the rotation generator
$\hat{p}_\varphi=\hat{M}_{12}$
under the CBT described in the previous section, and consider a complex   linear 
combination of the classical free particle symmetry generators 
$2iD_{0}+gp_\varphi$, where $g$ is a real parameter,
to produce the associated system
as a generalization of the isotropic harmonic oscillator that
is obtained  at $g=0$.
One has
$2iD_0+gp_\varphi=x_j\Delta_{jk}p_k$, where 
$\Delta_{jk}=i\delta_{jk}+g\epsilon_{jk}$ is
the complex tensor satisfying  the relations
\be\label{DDD}
\Delta_{jk}\Delta_{jl}=(g^2-1)\delta_{kl}\,,\qquad 
\det \Delta=g^2-1\,.
\ee
Based on (\ref{DDD}), 
one can expect
 that  
the  one-parametric  family of the quantum planar rotationally invariant  systems 
described by  
\be\label{QuantumHg}
\hat{H}_g=\hat{\mathfrak{S}}\omega (2i\hat{D}+g\hat{p}_\varphi)\hat{\mathfrak{S}}^{-1}=
\hat{H}_{\text{osc}}+g \omega \hat{p}_\varphi\,
\ee
should have essentially different physical
properties and symmetries 
 in the cases 
$g^2<1$ and $g^2>1$ separated
by the special parameter values $g=\pm 1$.
The symmetries  and the states of the quantum (and corresponding classical) system $\hat{H}_g$
has to be related by the CBT to those of the operator $2i\hat{D}+g\hat{p}_\varphi$ of the free
particle.

\vskip0.1cm
In terms of the ``circular" ladder operators, 
\begin{eqnarray}
&
\hat{b}_1^-= \frac{1}{\sqrt{2}}(\hat{a}_1^--i\hat{a}_2^-)\,, 
\quad \hat{b}_1^+=(\hat{b}_1^-)^\dagger\,,\quad
\hat{b}_2^-= \frac{1}{\sqrt{2}}(\hat{a}_{1}^-+i\hat{a}_{2}^-)
\,, \quad \hat{b}_2^+=(\hat{b}_2^-)^\dagger\,,\label{barel}
&
\end{eqnarray}
being unitary transformation of 
$\hat{a}_i^\pm$
and satisfying relations 
$[\hat{b}_i^\pm,\hat{b}_j^\pm]=0$,  
$[\hat{b}_i^-,\hat{b}_j^+]=\delta_{ij}$, 
the Hamiltonian  $\hat{H}_g$ takes the form 
 \be
 \label{QmrotHg0}
 \hat{H}_g= \hbar\omega(\ell_1\hat{b}_1^+\hat{b}_1^-  + \ell_2\hat{b}_2^+\hat{b}_2^- +1)\,,\qquad
 \ell_1=1+g\,,\qquad \ell_2=1-g\,.
 \ee
 This expression for  $ \hat{H}_g$ reminds 
 us  the  AHO Hamiltonian, 
 but presented here in a 
  rotationally invariant form. 
  The angular momentum operator 
  $\hat{p}_\varphi=\epsilon_{ij}\hat{x}_j\hat{p}_{j}=-i\hbar \epsilon_{ij}\hat{a}^+_i\hat{a}^-_j$
  is represented as
   \be
 \label{pphi}
  \hat{p}_\varphi=\hbar(\hat{b}_1^+\hat{b}_1^- - \hat{b}_2^+\hat{b}_2^-)\,,
  \ee
 and it commutes with  $\hat{H}_g$, $[\hat{H}_g,  \hat{p}_\varphi]=0$.
 
 The one-parameter family (\ref{QmrotHg0}) 
 of the ERIHOs is interesting 
as 
 it interpolates between
 different types of mechanical systems depending on the value of the real parameter $ g $:
 \begin{itemize}
\item[1] When $g=0$, we have the planar  isotropic harmonic oscillator.
\item[2]  
For
 $g=\pm 1$, the system corresponds to the Landau problem
 of a particle of charge $q$ in magnetic field  $B_3=\epsilon_{ij}\partial_i A_j^\pm=\mp B$
given by  the two-dimensional vector potential in symmetric gauge
\begin{eqnarray}
&
\label{Ag}
A_i^\pm=\pm \frac{1}{2}{B}\epsilon_{ij}x_j
\quad \Rightarrow\quad
\omega=\omega_B\equiv
\frac{qB}{2mc}\,,
&
\end{eqnarray}
where we assume $qB> 0$.
\item[3.] The case $|g|<1$ looks like the Euclidean AHO  
with different frequencies 
 $\omega_1\neq  \omega_2$,
$\omega_i=\ell_i\omega$. 
When $|g|>1$ we have 
 instead the form of the Minkowskian  AHO
 with frequencies  of two different signs.
 Notice that the family with $|g|>1$ 
resembles  the  Pais-Uhlenbeck  oscillator   \cite{PU1}
{}\footnote{Pais-Uhlenbeck  oscillator attracted recently considerable  attention
in relation to the $\mathcal{PT}$-symmetry, see  Refs. \cite{BenMan,Smilga,MostPU}.
}.

\item[4.] In the limit $g\rightarrow \infty$, one 
has $g^{-1}\hat{H}_g\rightarrow\omega \hat{p}_\varphi$.
In terms of the  operators $\hat{b}_i^\pm$ this corresponds to the isotropic 
Minkowskian 
oscillator, see Eq. (\ref{pphi}).  
\end{itemize}
The first two cases were analysed in the light of the CBT in \cite{InzPlyWipf1}. In this 
section we investigate 
the 
properties
of the ERIHO (\ref{QuantumHg})
in the general case, at the classical 
 and quantum levels.  This will allow us
 to reveal rather
 nontrivial relations between dynamics and symmetries 
 of the systems with different values of the parameter $g$.
 In particular, between those corresponding to the isotropic 
Minkowskian  oscillator  case, $g^2=\infty$, on the one hand, and 
the cases of the   isotropic  harmonic oscillator, $g=0$, and  Landau problem, $g^2=1$.

\subsection{Classical picture}\label{ClasPic}

Let us  
consider the classical system described by the Hamiltonian 
\be
\label{gHamil}
H_{g}=H_{\text{osc}}+ g\omega  p_\varphi\,,\qquad
H_{\text{osc}}=\frac{1}{2m}p_ip_i+\frac{1}{2}m\omega^2 x_ix_i\,,
\ee 
being the  classical analog of (\ref{QuantumHg}). As in the quantum case, 
this Hamiltonian 
arises 
 by applying the 
classical CBT to the complex linear combination $\omega(2iD+gp_\varphi)$
of the symmetry generators of the free particle system.

\begin{figure}[H]
\begin{center}
\begin{subfigure}[c]{0.28\linewidth}
\includegraphics[scale=0.48]{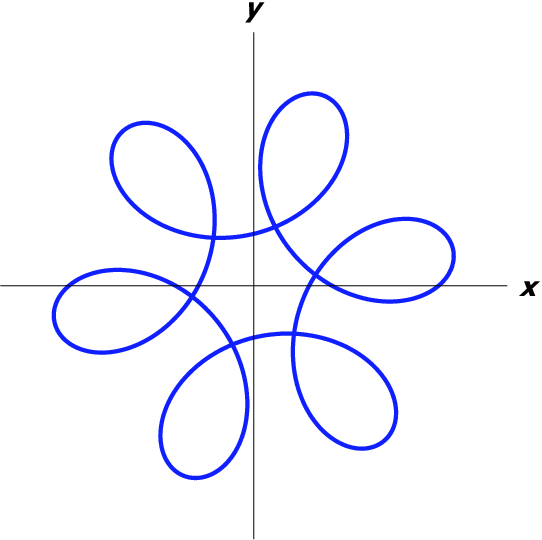}
\caption{\small{$g=2/3$, $R_1<R_2$}}
\end{subfigure}
\begin{subfigure}[c]{0.28\linewidth}
\includegraphics[scale=0.48]{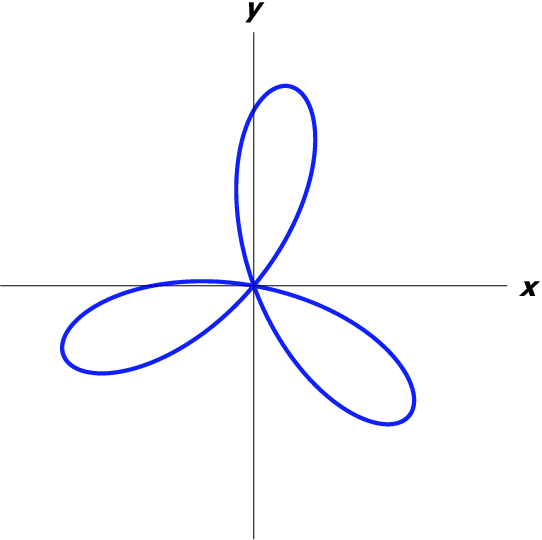}
\caption{\small{$g=1/3$, $R_1=R_2$}}
\end{subfigure}
\begin{subfigure}[c]{0.28\linewidth}
\includegraphics[scale=0.48]{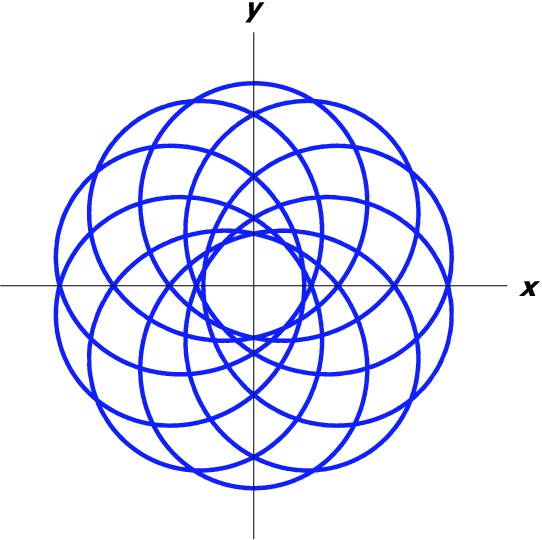}
\caption{\small{$g=4/5$, $R_1>R_2$}}
\end{subfigure}
\begin{subfigure}[c]{0.28\linewidth}
\includegraphics[scale=0.48]{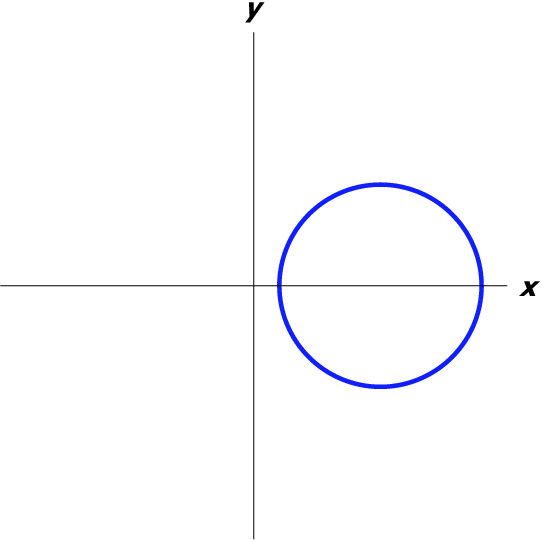}
\caption{\small{$g=1$, $R_1<R_2$}}
\end{subfigure}
\begin{subfigure}[c]{0.28\linewidth}
\includegraphics[scale=0.48]{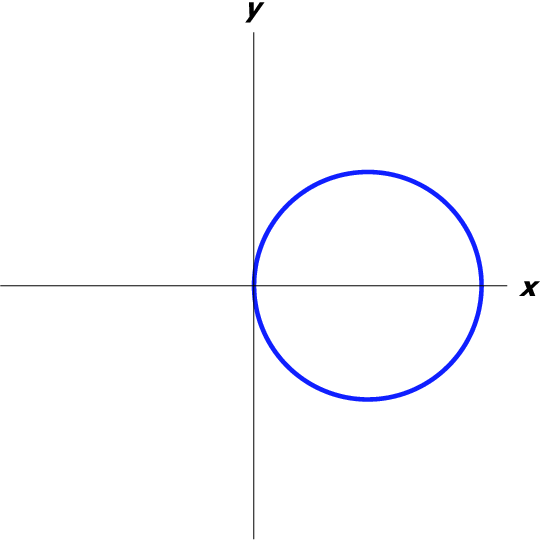}
\caption{\small{$g=1$, $R_1=R_2$}}
\end{subfigure}
\begin{subfigure}[c]{0.28\linewidth}
\includegraphics[scale=0.48]{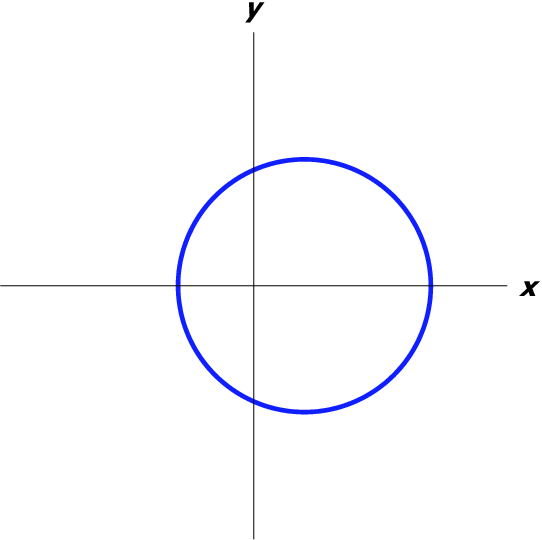}
\caption{\small{$g=1$, $R_1>R_2$}}
\label{Fig(f)}
\end{subfigure}
\begin{subfigure}[c]{0.28\linewidth}
\includegraphics[scale=0.48]{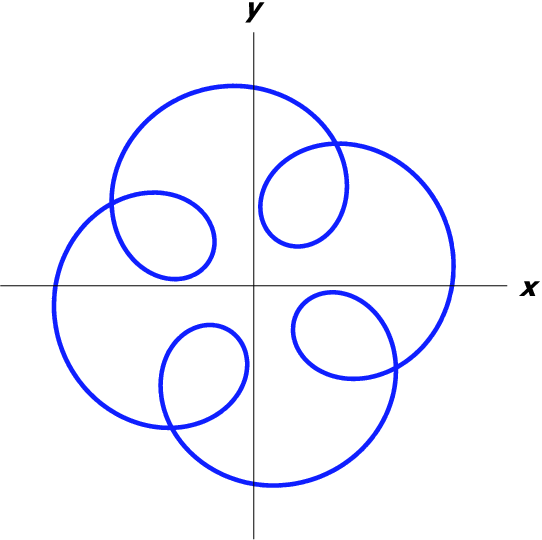}
\caption{\small{$g=3/2$, $R_1<R_2$}}
\end{subfigure}
\begin{subfigure}[c]{0.28\linewidth}
\includegraphics[scale=0.48]{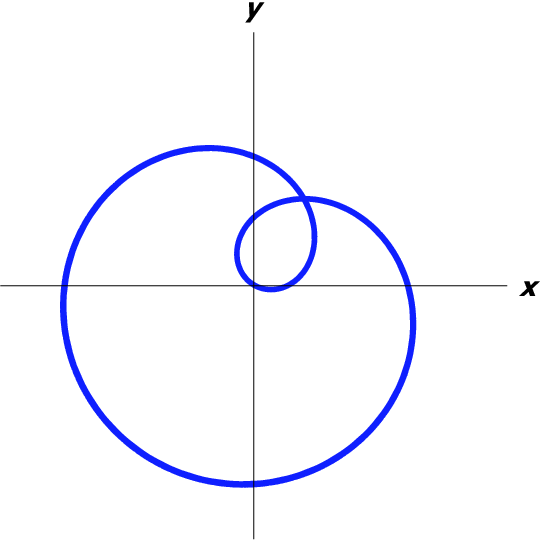}
\caption{\small{$g=3$, $R_1=R_2$}}
\end{subfigure}
\begin{subfigure}[c]{0.28\linewidth}
\includegraphics[scale=0.48]{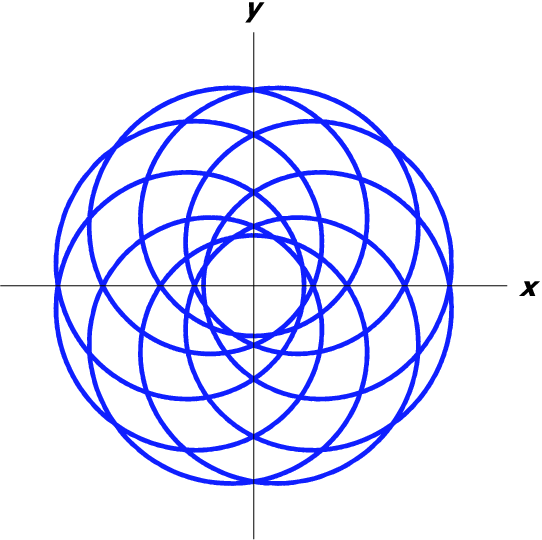}
\caption{\small{$g=5/4$, $R_1>R_2$}}
\end{subfigure}
\end{center}
\caption{\small{Trajectories for some rational values of $g$.
In cases b), e) and h),  $p_\varphi=0$ and trajectories pass through
 the origin.
 In cases (a), (d) and (g), $p_\varphi < 0$, while cases (c), (f) and (i) correspond to $p_\varphi > 0$.
}}
\label{Fig1}
\end{figure}

\begin{figure}[H]
\begin{center}
\begin{subfigure}[c]{0.28\linewidth}
\includegraphics[scale=0.37]{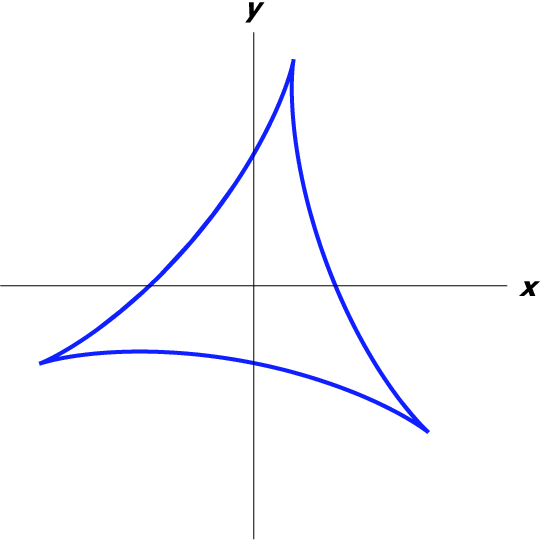}
\caption{\small{$g=1/3$, $R_2/R_1=2$}}
\end{subfigure}
\begin{subfigure}[c]{0.28\linewidth}
\includegraphics[scale=0.37]{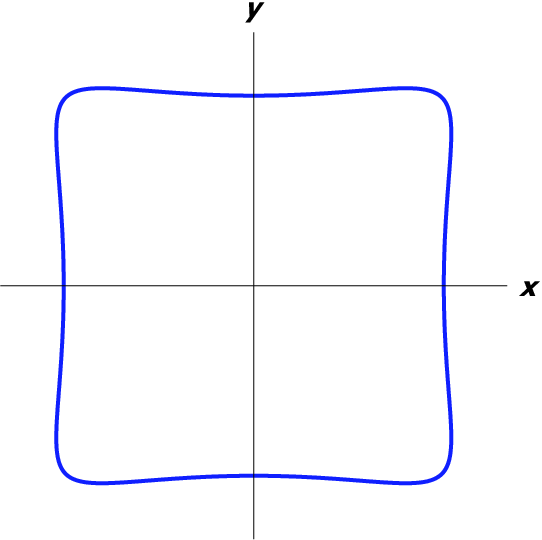}
\caption{\small{$g=1/2$, $R_2/R_1=6$}}
\end{subfigure}
\begin{subfigure}[c]{0.28\linewidth}
\includegraphics[scale=0.37]{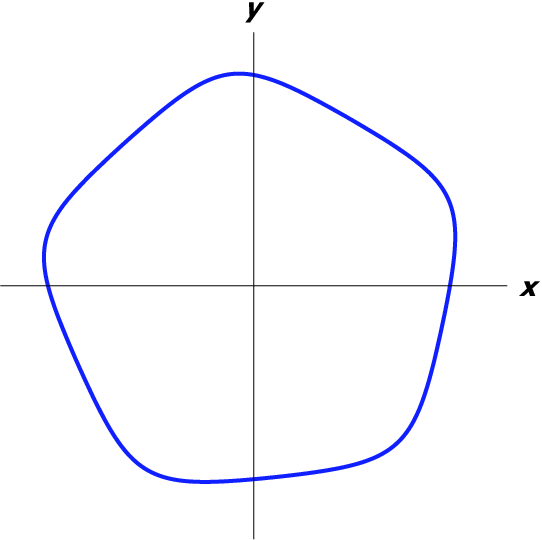}
\caption{\small{$g=3/5$, $R_2/R_1=20$}}
\end{subfigure}
\begin{subfigure}[c]{0.28\linewidth}
\includegraphics[scale=0.37]{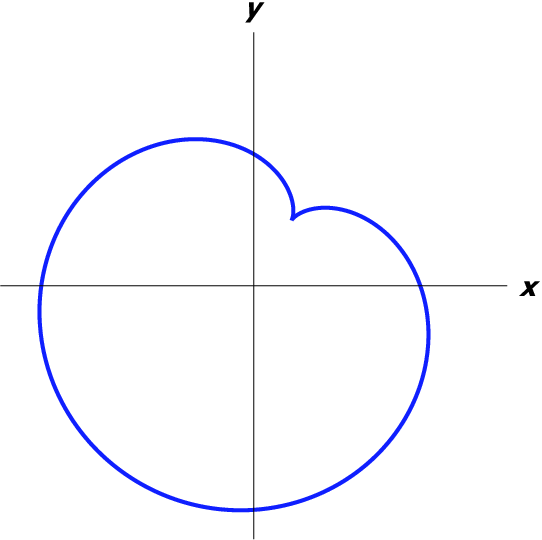}
\caption{\small{$g=3$, $R_2/R_1=2$}}
\end{subfigure}
\begin{subfigure}[c]{0.28\linewidth}
\includegraphics[scale=0.37]{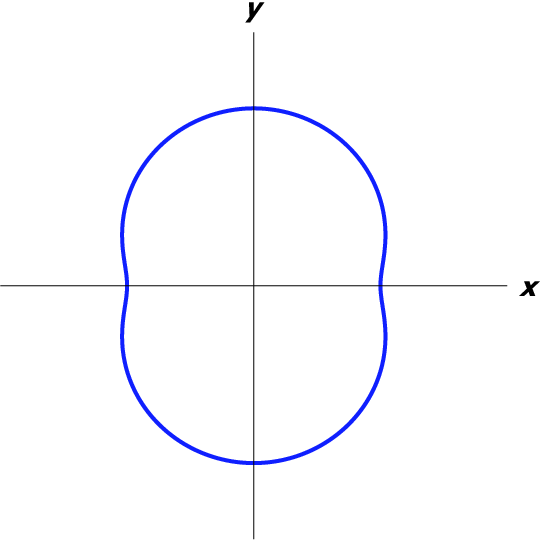}
\caption{\small{$g=2$, $R_2/R_1=6$}}
\end{subfigure}
\begin{subfigure}[c]{0.28\linewidth}
\includegraphics[scale=0.37]{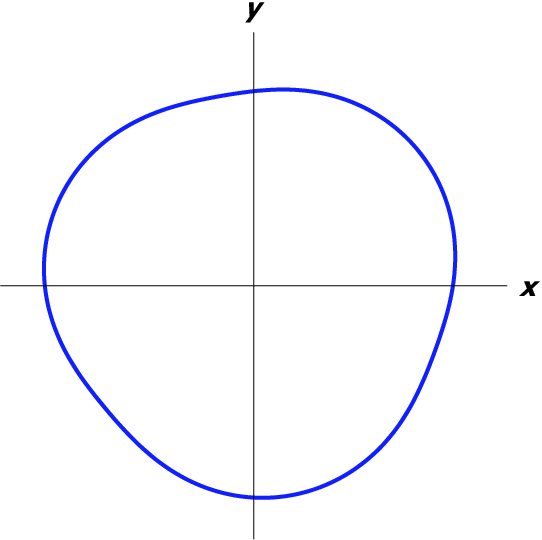}
\caption{\small{$g=5/3$, $R_2/R_1=20$}}
\end{subfigure}
\end{center}
\caption{\small{Trajectories for some rational values of $g$ 
and $R_1/R_2$. ``Dual" figures  (a)  and (d), see below,  correspond to a general 
 case
$R_1|\ell_1|=R_2|\ell_2|$ of the trajectories with cusps,
in which velocity turns into zero.
 In dual cases (b)-(e) and (c)-(f) the indicated equality is violated, 
 and corresponding trajectories are smooth.
}}
\label{Fig2}
\end{figure}

In terms of the classical analogues of 
circular ladder operators 
\begin{eqnarray}
\label{gen1}
&
b_1^-= \frac{1}{\sqrt{2}}(a_1^--ia_2^-)\,, \quad b_1^+=(b_1^-)^*\,,
\quad
b_2^-= \frac{1}{\sqrt{2}}(a_1^-+ia_2^-)\,,\quad b_2^+=(b_2^-)^*\,,
&\\&
a_i^\pm=\sqrt{\frac{m\omega}{2}}\left(\,x_i \mp \frac{i}{m\omega}\,p_i\right)\,,
\label{gen1+}&
\end{eqnarray}
Hamiltonian (\ref{gHamil}) takes the form 
\begin{eqnarray}
\label{gHamilbb+}
H_g= \omega\left(\ell_1 b_1^+b_1^-+\ell_2b_2^+b_2^- \right)\,,\qquad
\ell_1=1+g\,,\qquad \ell_2=1-g\,.
\end{eqnarray}
The equations of motion and their solutions are 
\be
\label{E.Q.O.M}
\dot{b}_i^\pm=\{b_i^\pm,H_g\}=\pm i\omega \ell_i b_i^\pm\quad
\Rightarrow \quad b_i^\pm(t)= e^{\pm i\omega \ell_i t }b_i^\pm(0):= b_i^\pm
\,.
\ee
Using (\ref{gen1}) and (\ref{gen1+}),  we 
have 
$\sqrt{m\omega}(x_1+ix_2)=b^+_1+b^-_2$, 
and 
find  
the trajectory of the particle, 
\be\label{traj}
z(t)=x_1(t)+ix_2(t)=R_1e^{i\gamma_1}e^{i\omega \ell_1 t}+R_2e^{-i\gamma_2}e^{-i\omega \ell_2 t}\,,
\ee
where  $R_i\geq 0$ and $\gamma_i\in \R$ are the  integration constants.
The exponents in (\ref{traj}) evolve in opposite,   
clockwise and counterclockwise, directions in the case $g^2<1$,
while for  $g^2>1$ they evolve in one of the two directions
 depending  on the sign of $g$. 
At $g^2=1$ one of the frequencies $\omega_i=\ell_i\omega$
 vanishes and (\ref{traj}) describes a closed
circular trajectory. At $g=+1$ ($g=-1$), $\omega_2=0$  ($\omega_1=0$),
and
the orbit is a circumference of radius $R_1$ ($R_2$) centered at 
$(X_1,X_2)$ with
$Z=X_1+iX_2=R_2e^{-i\gamma_2}$ ($Z=R_1e^{i\gamma_1}$).
In general, the trajectories will be closed for arbitrary choice of
the initial data (integration constants) 
 iff the condition 
$ \ell_1 / \ell_2 = q_2 / q_1 $ 
with 
$q_1, q_2 \in \Z $ 
is  fulfilled,  
that implies rational values for the parameter  
$g=(q_2-q_1)/(q_1+q_2)$. 
Some  trajectories for rational values of $g$ are shown in 
 Figs. \ref{Fig1} and \ref{Fig2}.
 \vskip0.5cm

 Rescaling the frequency parameter, $\omega\rightarrow \omega/|g|$, and taking  limit
 $|g|\rightarrow \infty$, Hamiltonian (\ref{gHamilbb+}) reduces to
 the Hamiltonian of the isotropic Minkowskian oscillator
 \be\label{sinfty}
 H_\infty= \epsilon_\infty \omega p_\varphi= \epsilon_\infty \omega
 (b^+_1b^-_1- b^+_2b^-_2)\,,
 \ee
 where $\epsilon_\infty=\pm1$   
  for $g\rightarrow \pm\infty$. 
 Solution to equations of motion for  system (\ref{sinfty}) are obtained from
 (\ref{traj})
 by the same procedure $\omega\rightarrow \omega/|g|$, 
 $|g|\rightarrow \infty$, 
 \be\label{traj+}
z(t)=x(t)+iy(t)=R_1e^{i(\omega  t+\gamma_1)}+R_2e^{i(\omega  t-\gamma_2)}\,.
\ee
For the sake of definiteness we assume here  
$\epsilon_\infty=+1$.
 Eq. (\ref{traj+}) describes a circular trajectory 
 centered at the origin,  for which   the squared radius is given by
 \be\label{MinTraj}
 x^2(t)+y^2(t)=R_1^2+R_2^2+2R_1R_2\cos(\gamma_1+\gamma_2)\,,
 \ee
 and so, 
 $(R_1-R_2)^2\leq x^2(t)+y^2(t)\leq (R_1+R_2)^2$.
 From the viewpoint of dynamics,
 the case of the isotropic Minkowskian oscillator
 is similar, on the one hand,  to the case of Euclidean isotropic oscillator $(g=0)$,
 whose trajectories also are  centered at the origin, but which are
 ellipses that reduce to circular trajectories  only for
  particular choice of the  initial data.
 On the other hand, 
 the isotropic Minkowskian oscillator 
  is similar to the case of Landau problem ($g^2=1$),
 where trajectories are circular, but which are centered  at the origin
 only for a  particular choice of the  
 initial data.

 The explicitly depending on time complex phase space functions 
 \be\label{betaint}
 \mathcal{\beta}^\pm_j=b_j^\pm  e^{\mp i\omega \ell_j t }\,,\quad
 j=1,2\,,
 \ee
that correspond to  the integration constants 
$b_j^\pm(0)=R_je^{\pm i\gamma_j}$, are the \emph{dynamical}
integrals of motion, which generate the two-dimensional Heisenberg algebra. 
In the case $g=+1$  ($g=-1$),  one pair of them 
transforms into the  \emph{true}, not depending explicitly on time 
mutually conjugate complex integrals of 
motion of the Landau problem. Being multiplied by $\frac{1}{\sqrt{m\omega}}$, they correspond 
to the coordinates $(X_1,X_2)$ 
of the center of the circular orbit 
having nonzero Poisson brackets  $\{X_1,X_2\}=\frac{g}{2m\omega}$, $g=\pm 1$.
Using these, dynamical in the case  $g^2\neq 1$
 integrals (\ref{betaint}), one can construct 
ten quadratic 
integrals 
\begin{eqnarray}
\label{sl2Rosc}
&\mathcal{J}_{\pm}= 
e^{\mp 2i\omega t} b_1^\pm b_2^\pm\,,\qquad
\mathcal{J}_0=\frac{1}{2}( b_1^+ b_1^-+ b_2^+ b_2^-)=\frac{1}{2\omega} H_{\text{osc}}\,,&\\&
\mathcal{L}_2=\frac{1}{2}(b_1^+b_1^--b_2^+b_2^-)=\frac{1}{2}p_\varphi
\,,\qquad
\mathcal{L}_\pm=
 e^{\mp 2i\omega  g t}b_1^\pm b_2^\mp\,,&\label{u1osc}\\
&\qquad
\mathcal{B}_1^\pm=e^{\mp 2i\omega\ell_1 t}(b_1^\pm)^{2}\,,\qquad
\mathcal{B}_2^\pm=e^{\mp 2i\omega\ell_2 t}(b_2^\pm)^{2}\,.
&\label{BB}
\end{eqnarray}
In this set, only $\mathcal{J}_0$ and $\mathcal{L}_2$ are the 
true integrals  for general case of $g$ 
since
$H_g=2\omega (\mathcal{J}_0+g\mathcal{L}_2)$ and $\{\mathcal{J}_0,\mathcal{L}_2\}=0$. 
Here,  the quantities  (\ref{sl2Rosc}) and  (\ref{u1osc})
generate, respectively,  the $\mathfrak{sl}(2,\R)$ 
and the $\mathfrak{su}(2)$ algebras.  
All the quadratic integrals (\ref{sl2Rosc}), (\ref{u1osc}) and 
(\ref{BB})
generate  
the $\mathfrak{sp}(4,\R)$ algebra  with the following nonzero 
Poisson brackets, 
\begin{eqnarray}
\label{Sl2Ros}
&
\{\mathcal{J}_0,\mathcal{J}_\pm\}=\mp i \mathcal{J}_\pm\,,\qquad
\{\mathcal{J}_-,\mathcal{J}_+\}=-2i\mathcal{J}_{0}\,, &\\&
\label{Su(2)OS}
\{\mathcal{L}_2,\mathcal{L}_\pm\}=\mp i\mathcal{L}_\pm\,,\qquad
\{\mathcal{L}_+,\mathcal{L}_-\}=-2i \mathcal{L}_2\,, &\\&
\label{sp4+1}
\{\mathcal{J}_\pm,\mathcal{L}_\mp\}=\pm i\mathcal{B}_2^\pm\,,\qquad 
\{\mathcal{J}_\pm,\mathcal{L}_\pm\}=\pm i\mathcal{B}_1^\pm\,, &
\\&
\label{sp4+2}
\{\mathcal{J}_0, \mathcal{B}_j^\pm\}=\mp i\mathcal{B}_j^\pm\,,\qquad
\{\mathcal{J}_\mp,\mathcal{B}_2^\pm \}=\mp 2i\mathcal{L}_{\mp}\,,\qquad 
\{\mathcal{J}_\mp,\mathcal{B}_1^\pm \}=\mp 2i\mathcal{L}_{\pm}\,, &\\&
\label{sp4+3}
\{\mathcal{L}_{2}, \mathcal{B}_j^\pm \}= \pm (-1)^j  i \mathcal{B}_j^\pm \,,\qquad
\{\mathcal{L}_{\pm}, \mathcal{B}_1^\mp \}=\pm 2i \mathcal{J}_\mp\,,\qquad
\{\mathcal{L}_{\pm}, \mathcal{B}_2^\pm \}=\mp 2i \mathcal{J}_\pm\,,&\\&
\{\mathcal{B}_1^-,\mathcal{B}_1^+\}=-4i\left(\mathcal{J}_0+\mathcal{L}_2\right)\,,\qquad
\{\mathcal{B}_2^-,\mathcal{B}_2^+\}=-4i\left(\mathcal{J}_0-\mathcal{L}_2\right)\,.
\label{Sl2Ros+}
 &
\end{eqnarray}
By taking repeatedly the Poisson brackets
of any of five sets of four integrals
$( \mathcal{J}_\pm, \mathcal{L}_\pm)$,
$( \mathcal{J}_\pm, \mathcal{B}_1^\pm)$,
$( \mathcal{J}_\pm, \mathcal{B}_2^\pm)$,
$( \mathcal{L}_\pm, \mathcal{B}_1^\pm)$ and 
$( \mathcal{L}_\pm, \mathcal{B}_2^\pm)$,
all the $\mathfrak{sp}(4,\R)$ algebra  is produced.
On the other hand, we notice that the algebra 
contains the $\mathfrak{su}(2)\oplus \mathfrak{u}(1)\cong \mathfrak{u}(2)$ subalgebra generated 
by the set of integrals 
\be\label{S1}
(S1):\quad ( \mathcal{L}_2, \mathcal{L}_\pm)\oplus  \mathcal{J}_0\,,
\qquad
\mathcal{L}_2^2+\mathcal{L}_+\mathcal{L}_-=\mathcal{J}_0^2\,,
\ee
where the second equality 
corresponds to relation between the Casimir element of the $\mathfrak{su}(2)$ 
subalgebra and central element $\mathcal{J}_0$ of 
the
 $ \mathfrak{u}(2)$.

The $\mathfrak{sp}(4,\R)$ algebra  also contains  three copies of 
the Lie algebra 
$\mathfrak{sl}(2,\R)\oplus \mathfrak{u}(1)\cong \mathfrak{gl}(2,\R)$ 
generated by any of the three  sets of integrals
\begin{eqnarray}\label{S2}
&(S2):\quad  (\mathcal{J}_0, \mathcal{J}_\pm)\oplus  \mathcal{L}_2\,,\qquad
-\mathcal{J}_0^2+\mathcal{J}_+\mathcal{J}_-=-\mathcal{L}_2^2\,,&
\\
&
 (S3):\quad ( \frac{1}{2}\left(\mathcal{J}_0- \mathcal{L}_2),\frac{1}{2} 
 \mathcal{B}_2^\pm\right)\oplus  (\mathcal{J}_0+ \mathcal{L}_2)\,,\quad
 -\left(\frac{1}{2}\left(\mathcal{J}_0- \mathcal{L}_2\right)\right)^2+
 \frac{1}{2}\mathcal{B}_2^+ \frac{1}{2} \mathcal{B}_2^-=0\,,\label{S3}&\\
 &
 (S4): \quad \left( \frac{1}{2}(\mathcal{J}_0+\mathcal{L}_2),\frac{1}{2} 
 \mathcal{B}_1^\pm\right)\oplus  (\mathcal{J}_0- \mathcal{L}_2)\,,
 \quad
 -\left(\frac{1}{2}\left(\mathcal{J}_0+\mathcal{L}_2\right)\right)^2+
 \frac{1}{2}\mathcal{B}_1^+\frac{1}{2}\mathcal{B}_1^-=0\,.
 \label{S4}&
 \end{eqnarray}
Here, analogously to (\ref{S1}), we indicated  the values taken by
 the  Casimir element of the $\mathfrak{sl}(2,\R)$ subalgebra
 in each copy of  $\mathfrak{gl}(2,\R)$.
  
 In the case $g=0$, besides $\mathcal{J}_0$ and $\mathcal{L}_2$,
 the system has two additional true integrals of motion
 $\mathcal{L}_\pm$ 
 not depending explicitly on time,
 and so, 
 the  $\mathfrak{u}(2)$  subalgebra 
 is the true symmetry of the isotropic harmonic oscillator
 $H_{\text{osc}}$. For it,   the $\mathfrak{sl}(2,\R)$ part of the 
 (S2) subalgebra corresponds to 
 the dynamical conformal symmetry.
 
For Landau problem with $g=1$, 
the subalgebra  (S3) corresponds to the true symmetry 
of the system, while its dynamical conformal 
symmetry is generated by the $\mathfrak{sl}(2,\R)$  part
of the (S4) subalgebra. 
 In the case of Landau problem with $g=-1$,
 the true symmetry corresponds to subagebra (S4),
 and its dynamical conformal 
symmetry is generated 
 by the  $\mathfrak{sl}(2,\R)$ part of the subalgebra (S3).

In the case of the isotropic Minkowskian oscillator (obtained by 
$\omega\rightarrow \omega/|g|$, $|g|\rightarrow \infty$),
the subalgebra (S2) with its $\mathfrak{sl}(2,\R)$
part corresponds to the true symmetry,
while the $\mathfrak{su}(2)$ part of the 
 (S1) subalgebra is its dynamical conformal symmetry. 
 We have here a kind of transmutation of 
 the true symmetry  into dynamical conformal symmetry and vise versa 
  when we pass over from 
 $g=0$   
 to the $|g|=\infty$ case. 
  We return to this point below in Section \ref{Landau}.
   Analogous phenomenon 
  of transmutation of symmetries takes place 
  for the Landau problem corresponding to the cases of
   $g=+1$ and $g=-1$. However, one notes that 
   in the Landau
   problem in both cases the true and dynamical conformal 
   symmetries correspond to  different realizations of the same 
    $\mathfrak{gl}(2,\R)$ algebra.
   
   In the case of the isotropic Euclidean  oscillator ($g=0)$,
   the integrals $\mathcal{J}_0=\frac{1}{2\omega}H_{\text{osc}}$
   and $\mathcal{L}_2=\frac{1}{2}p_\varphi$ define 
   the major and minor semi-axes of the elliptic orbit,
   while the  pair of the integrals 
   $\mathcal{L}_\pm$ defines its orientation in the plane 
   via their phase 
   $(\gamma_1-\gamma_2)$, see Eq. (\ref{traj}).
   The modulus of     $\mathcal{L}_\pm$ is fixed by the
   integrals $\mathcal{J}_0$ and $\mathcal{L}_2$
   via the $\mathfrak{su}(2)$  Casimir value, see  Eq. 
   (\ref{S1}).
   
   In the case of the  isotropic  Minkowskian oscillator,
  the radius of  its circular orbit centered at the origin is defined 
  by the three parameters $R_1$, $R_2$ and $(\gamma_1+\gamma_2)$,
  see Eq. (\ref{traj+}), which are fixed, again,  by 
  the integrals $\mathcal{L}_2=\frac{1}{2\omega}H_\infty$, 
  $\mathcal{J}_0$,  and by the phase of the true integrals 
  $\mathcal{J}_\pm$ of the system. The modulus 
  of  $\mathcal{J}_\pm$ is fixed, in turn,  
  by the
   integrals $\mathcal{J}_0$ and $\mathcal{L}_2$
   via the  $\mathfrak{sl}(2,\R)$ Casimir value, see Eq. 
   (\ref{S2}).
   
   In the Landau problem, the radius of the circular orbit is defined
   by the corresponding Hamiltonian $H_{g=1}=2\omega (\mathcal{J}_0+\mathcal{L}_2)$
   ($H_{g=-1}=2\omega (\mathcal{J}_0-\mathcal{L}_2)$),
   while its center is given by the $\mathfrak{sl}(2,\R)$ 
   generators
   $\mathcal{B}_2^\pm$ ($\mathcal{B}_1^\pm$)
   (being the squares of the corresponding linear  
   integrals $\beta^\pm_2$ ($\beta^\pm_1$)). 
   Zero values of the $\mathfrak{sl}(2,\R)$  classical Casimirs
   in Eqs. (\ref{S3}) and (\ref{S4}) reflect 
   the fact that in the Landau problem 
   dynamics is effectively   governed  by the 
   corresponding one-dimensional harmonic oscillator
  Hamiltonians  $H_{g=1}=\omega b^+_1b^-_1$ ($H_{g=-1}=b^+_2 b^-_2$)
  \cite{InzPlyWipf1}.

 The closed character of the trajectories for rational values of the parameter 
 $g$ different from the already discussed cases
 of  $g= 0,\pm 1,\pm \infty$  
  indicates that some additional true integrals  of motion also have to appear 
 in the corresponding systems.
 Such integrals,  however, are of higher order and, as we will see,
 produce  non-linear deformations
 of the  $\mathfrak{u}(2)$  and 
 $\mathfrak{gl}(2,\R)$
symmetries
  in the cases 
 of $g^2<1$ and $g^2>1$, respectively.
They can be found by taking the products of dynamical integrals $\beta^+_j$ and
$\beta^-_j$  so that   the time-dependent 
exponential factors in them will be cancelled.
 For this, consider  the dynamical integrals 
\begin{eqnarray}
&\label{higherorderA}
\mathcal{L}_{j_1,j_2}^{+}=(\beta_1^+)^{j_1}(\beta_2^-)^{j_2}= e^{-i\omega (j_1\ell_1-j_2\ell_2 )t }(b_1^+)^{j_1} (b_2^-)^{j_2}\,.
\qquad \mathcal{L}_{j_1,j_2}^{-}=(\mathcal{L}_{j_1,j_2}^{+})^*\,,
&
\end{eqnarray}
where, in principle, the indexes  $j_1$, $j_2$ can take any 
non-negative  integer values. 
In order (\ref{higherorderA}) would be true integrals for the system $H_g$, 
$\{H_g,\mathcal{L}_{j_1,j_2}^{\pm}\}=0$, 
there should exist the  exponents  $j_i=s_i$
that obey the relation
\be
\label{condition 1}
s_1\ell_1-s_2\ell_2=0\,.
\ee 
The condition (\ref{condition 1}) is satisfied iff
$g=(s_2-s_1)/(s_1+s_2)$. 
The  positive integer numbers $s_1$ and $s_2$  can be chosen 
in such a way that the fraction is irreducible, 
that we will imply in what follows.
So, two additional  higher order  true integrals 
$\mathcal{L}_{s_1,s_2}^\pm$ of the indicated form exist 
for rational values of $g$ with $|g|<1$. 

The  time-independent  integrals $\mathcal{L}_{s_1,s_2}^\pm$ 
are eigenstates of the true integrals ${\mathcal{J}}_0$, $\mathcal{L}_2$ 
in the sense of Poisson brackets, while 
dynamical integrals 
${\mathcal{J}}_\pm$, 
change their indexes  by transforming them into 
dynamical integrals:
\begin{eqnarray}\label{GenA0}
&
\{{\mathcal{L}}_2,\mathcal{L}_{s_1,s_2}^\pm\}=\mp \frac{i}{2}(s_1+s_2)\mathcal{L}_{s_1,s_2}^\pm\,,\qquad
\{{\mathcal{J}}_0,\mathcal{L}_{s_1,s_2}^\pm\}=\mp \frac{i}{2}(s_1-s_2) \mathcal{L}_{s_1,s_2}^\pm\,,
&\\&
\label{GenA}
\{\mathcal{J}_\pm,\mathcal{L}_{s_1,s_2}^\pm\}=\pm is_2 \mathcal{L}_{s_1+1,s_2-1}^\pm\,,\qquad
\{\mathcal{J}_\mp,\mathcal{L}_{s_1,s_2}^\pm\}=\mp is_1 \mathcal{L}_{s_1-1,s_2+1}^\pm\,.&
\end{eqnarray}
Taking Poisson brackets of ${\mathcal{J}}_0$, $\mathcal{L}_2$ and  
${\mathcal{J}}_\pm$ with the generated integrals 
$\mathcal{L}_{s_1+1,s_2-1}^\pm$ and $\mathcal{L}_{s_1-1,s_2+1}^\pm$,
we continue the process,  producing in this way
 the finite set of the integrals 
\be\label{Lpmss}
 \mathcal{L}_{s_1 + s_2,0}^{\pm},\,\ldots,\mathcal{L}_{s_1,s_2}^\pm,\,\ldots
 \,\mathcal{L}_{0,s_1 + s_2}^{\pm} \,,
 \ee
 in which only $\mathcal{L}_{s_1,s_2}^\pm$ are the true, not depending explicitly on time,   integrals,
 while the rest are dynamical, time-dependent higher order integrals.
 Using the Jacobi identity, one can show that 
 the phase space functions  $\{\mathcal{L}_{n_1,n_2}^-,\mathcal{L}_{r_1,r_2}^+\}$ with  $n_1+n_2=r_1+r_2=s_1+s_2$ 
 Poisson commute with $\mathcal{L}_2$, and therefore, this bracket 
must be a function  of  the  $\mathfrak{sl}(2,\R)$ generators   (\ref{sl2Rosc}) 
and the angular momentum. 
Moreover,  one has 
\be
\{\{\mathcal{L}_{s_1,s_2}^-\,,\mathcal{L}_{s_1,s_2}^+\},\mathcal{J}_0\}=0\quad 
\Rightarrow\quad 
\{\mathcal{L}_{s_1,s_2}^-,\mathcal{L}_{s_1,s_2}^+\}=F_g(H_g,\mathcal{L}_2)\,,
\ee
where $F_g$ is a 
 polynomial function of $H_g$ and  $\mathcal{L}_2$.
In conclusion, the set 
(${\mathcal{J}}_0$, ${\mathcal{J}}_\pm$, $\mathcal{L}_2$, $\mathcal{L}_{k,s_1+s_2-k}^\pm$),
 where 
$k=0,\ldots, s_1+s_2$, produces a finite dimensional non-linear algebra,
in which the non-linear subalgebra generated by true integrals 
$H_g$, $\mathcal{L}_2$ and $\mathcal{L}_{s_1,s_2}^\pm$ 
corresponds to a 
deformation of   $\mathfrak{u}(2)$.
 
The already discussed in detail cases of the isotropic Euclidean 
oscillator and Landau problems with their  $\mathfrak{sp}(4,\R)$
Lie algebra  can also be included in the 
described structure. The case  
$s_1=s_2=1$ reproduces here the  case $g=0$, for which  
 the integrals $\mathcal{L}_2$  and 
 $\mathcal{L}_{1,1}^\pm=\mathcal{L}_{\pm}$, 
generate the  $\mathfrak{su}(2)$ hidden symmetry of the system.
On the other hand $s_1=0$ and  $s_2=2$ ($s_1=2$ and $s_2=0$) yields 
$g=1$ ($g=-1$),  for which $\mathcal{L}_{0,2}^\pm=2 \mathcal{B}^\pm_{2}$
($\mathcal{L}_{2,0}^\pm=2 \mathcal{B}^\pm_{1}$). 
Taking $s_1=0$ and $s_2=1$ ($s_1=1$ and $s_2=0$),
we reproduce 
the first order true integrals for the Landau problem with   $g=1$ ($g=-1$),
$\mathcal{L}_{0,1}^\pm=b_2^\mp$ ($\mathcal{L}_{1,0}^\pm=b_1^\mp$),
that generate translations of particle's coordinates $x_i$.

If instead of (\ref{higherorderB}) we consider the dynamical integrals 
\begin{eqnarray}
&
\label{higherorderB}
\mathcal{J}_{j_1,j_2}^+=(\beta_1^+)^{j_1}(\beta_2^+)^{j_2}= 
e^{-i\omega (j_1\ell_1+j_2\ell_2 )t }(b_1^+)^{j_1} (b_2^+)^{j_2}\,,\qquad
 \mathcal{J}_{j_1,j_2}^{-}=(\mathcal{J}_{j_1,j_2}^{+})^*\,,
&
\end{eqnarray}
one notes that 
the  $\mathcal{J}_{s_1,s_2}^\pm$ are the time-independent,  true integrals of motion if an only if 
the condition 
\be
\label{condition 2}
s_1\ell_1+s_2\ell_2=0\,,
\ee
is satisfied, i.e.  when $g=(s_2+s_1)/(s_2-s_1)$, that implies $|g|>1$.
In this case, 
we
consider the Poisson bracket  relations of 
$\mathcal{J}_{s_1,s_2}^{\pm}$  with the true integrals 
$\mathcal{J}_0$ and $\mathcal{L}_2$ and 
two other generators 
 $\mathcal{L}_\pm$ of the dynamical
$\mathfrak{su}(2)$ symmetry:
\begin{eqnarray}
&
\{\mathcal{J}_0,\mathcal{J}_{s_1,s_2}^\pm \}
=\mp\frac{i}{2}(s_1+s_2)\mathcal{J}_{s_1,s_2}^\pm\,,\qquad
\{\mathcal{L}_2,\mathcal{J}_{s_1,s_2}^\pm \}=\mp \frac{i}{2}(s_1-s_2)\mathcal{J}_{s_1,s_2}^\pm\,,&\\&
\{\mathcal{L}_\pm,\mathcal{J}_{s_1,s_2}^\pm\}=\mp is_2 \mathcal{J}_{s_1+1,s_2-1}^+\,,\qquad
\{\mathcal{L}_\mp,\mathcal{J}_{s_1,s_2}^\pm\}=\pm is_1 \mathcal{J}_{s_1-1,s_2+1}^+\label{genB}\,.&
\end{eqnarray} 
By the same reasoning as in the case of rational $g$  with 
$|g|<1$,  we generate  the set 
\be\label{Jpmss}
\mathcal{J}_{s_1+s_2,0}^\pm,\,\ldots,\,\mathcal{J}_{s_1,s_2}^\pm,\,
\ldots,\,
\mathcal{J}_{0,s_1+s_2}^\pm\,,
\ee
in which only $\mathcal{J}_{s_1,s_2}^\pm$ are the true integrals,
while the rest are the dynamical integrals of motion. 
They together with 
 $\mathcal{J}_0$, $\mathcal{L}_2$ and $\mathcal{L}_\pm$
 generate a finite non-linear algebra, in which 
   $\{\mathcal{J}_{j_1,j_2}^-,\mathcal{J}_{j_1,j_2}^+\}$ 
   is a polynomial  function of $H_g$ and $\mathcal{L}_2$ only.
Taking here $s_1=0$ and $s_2=2$ ($s_1=2$ and $s_2=0$),
we reproduce the quadratic integrals $\mathcal{J}_{0,2}^\pm=2\mathcal{B}^\pm_2$
($\mathcal{J}_{2,0}^\pm=2\mathcal{B}^\pm_1$)
of the Landau problem with $g=1$ ($g=-1$).
The values  $s_1=0$ and $s_2=1$ ($s_1=1$ and $s_2=0$)
provide us with the corresponding linear  integrals 
of the Landau problem with $g=1$ ($g=-1$).
On the other hand,  
setting formally  $s_1=s_2=1$,  
we reproduce the true 
integrals $\mathcal{J}_{1,1}^\pm=\mathcal{J}_\pm$ 
of the isotropic Minkowskian oscillator,
which together with $\mathcal{J}_0$ 
and $H_\infty=2\omega\mathcal{L}_2$ generate 
its Lie algebraic $\mathfrak{gl}(2,\R)$ symmetry. 
In the case of finite  rational  values of $g$ with $|g|>1$, 
the subalgebra generated by $\mathcal{J}_{s_1,s_2}^\pm$, $\mathcal{J}_{0}$
and $H_g$
is identified as a non-linear deformation of the $\mathfrak{gl}(2,\R)$.  

We do not consider here a rather complicated 
complete non-linear Poisson bracket 
algebraic structure generated by the true and dynamical integrals 
in the case of rational values of the parameter
$g$ different from the already discussed particular cases. 
We only note  that 
the repeated Poisson brackets of higher order 
dynamical integral 
$\mathcal{L}_{s_1+s_2,0}^{+}= \mathcal{J}_{s_1+s_2,0}^{+}$
 ($\mathcal{L}_{s_1+s_2,0}^{-}= \mathcal{J}_{s_1+s_2,0}^{-}$)
of the case  $g=(s_2-s_1)/(s_2+s_1):=g_<^{s_1,s_2}$, $|g_<^{s_1,s_2}|<1$, with quadratic 
dynamical integrals $\mathcal{L}_-$ 
($\mathcal{L}_+$)
generate all the set (\ref{Jpmss})
of the integrals that we have had in the case 
of $H_g$ with $g=(s_2+s_1)/(s_2-s_1):=g_>^{s_1,s_2}=1/g_<^{s_1,s_2}$,
$|g_>^{s_1,s_2}|>1$. 
All the integrals
(\ref{Jpmss}) are, however,  dynamical 
for the system $H_g$ with 
$g=g_<^{s_1,s_2}$.
Analogously, 
the repeated Poisson brackets of higher order 
dynamical integral 
$\mathcal{J}_{s_1+s_2,0}^{+}$ ($\mathcal{J}_{s_1+s_2,0}^{-}$)
of the case  
 $g=g_>^{s_1,s_2}$ 
 with quadratic 
dynamical integrals 
$\mathcal{L}_-$ 
($\mathcal{L}_+$)
generate all the set (\ref{Lpmss})
of the integrals that we have had in the case 
of $H_g$ with 
$g=g_<^{s_1,s_2}$. 
All integrals
(\ref{Lpmss}) are dynamical 
for the system $H_g$ with 
$g=g_>^{s_1,s_2}$.
From this point of view 
we also have a kind of transmutation 
of symmetries for the 
``dual" pairs of 
the systems with 
$g=g_<^{s_1,s_2}$
 and $g=g_>^{s_1,s_2}=1/g_<^{s_1,s_2}$,
 where the non-linearly 
 deformed  $\mathfrak{u}(2)$ 
 and     $\mathfrak{gl}(2,\R)$
 subalgebras generated by the sets 
 ($H_g,\mathcal{L}_{2},\mathcal{L}_{s_1,s_2}^{\pm}$) 
 and 
  ($H_g,\mathcal{J}_{0},\mathcal{J}_{s_1,s_2}^{\pm}$) 
change their role in the sense 
of the true and dynamical sub-symmetries.

\subsection{Quantum picture}
\label{SecRIAHOq}

Now, we return to the quantum system (\ref{QuantumHg}) to 
analyze,  in 
  the light of the CBT,  its spectrum, integrals associated with degeneracy of 
the energy levels,  and  
 spectrum generating ladder 
 operators.

To find
normalizable eigenstates and the spectrum of the system (\ref{QuantumHg}) by means 
of the CBT, we  exploit its  rotational invariance reflected,  particularly, 
in the form of classical solutions (\ref{traj}), and pass over from
Cartesian coordinates $x_i$ to the complex variable
$z=x_1+ix_2$,
 $\frac{\partial}{\partial z}=\frac{1}{2}(\frac{\partial}{\partial x_1}-i\frac{\partial}{\partial x_2})$.
 In terms of $z$ and $z^*$, the equation 
 for formal eigenfunctions  $\phi_{\lambda}(x_1,x_2)=\bra{x_1,x_2}\ket{\lambda}$
 of the non-Hermitian operator $2i\hat{D}+g\hat{p}_\varphi$
 takes the form
\begin{eqnarray}\label{Dpeig}
&
(2i\hat{D}+g\hat{p}_\varphi)\phi_{\lambda}=\hbar\left( (1+g)z\frac{\partial}{\partial z}+
(1-g)z^{*}\frac{\partial}{\partial z^{*}}
 \right)\phi_{\lambda}=\lambda\phi_{\lambda}\,.
&
\end{eqnarray}
The well defined in $\R^2$ simultaneous eigenfunctions  of 
the mutually commuting operators $2i\hat{D}$ and $\hat{p}_\varphi$
are
 \be
 \label{JordanStates}
 \phi_{n_1,n_2}=z^{n_1}(z^*)^{n_2}\,,
 \ee 
 where $n_1$ and $n_2$ are non-negative integers.
They   satisfy relations 
  \begin{eqnarray}
 &
 \hat{p}_-\phi_{n_1,n_2}=-2i\hbar n_1 \phi_{n_1-1,n_2}\,,\qquad
 \hat{p}_+\phi_{n_1,n_2}=-2i\hbar n_2 \phi_{n_1,n_2-1}\,,
 &
 \\
 &
\hat{\xi}_+ \phi_{n_1,n_2}=m\phi_{n_1+1,n_2}\,,\qquad
\hat{\xi}_- \phi_{n_1,n_2}=m\phi_{n_1,n_2+1}\,, 
\end{eqnarray}
where $\hat{p}_\pm=\hat{p}_1\pm i\hat{p}_2$ and  
$\hat{\xi}_\pm=\hat{\xi}_1\pm i\hat{\xi}_2$.
Therefore,  free particle quadratic 
operators 
$\hat{H}$, $\hat{D}$,
$\hat{K}$ and $\hat{p}_\varphi$ act on  states 
(\ref{JordanStates}) as follows:
 \begin{eqnarray}
&
\hat{H}\phi_{n_1,n_2}=-\frac{2\hbar}{m}n_1n_2 \phi_{n_1-1,n_2-1}\,,\qquad
\hat{K}\phi_{n_1,n_2}=\frac{m}{2} \phi_{n_1+1,n_2+1}\,,\qquad
\label{HyKonphi}
&\\&
  2i\hat{D}\phi_{n_1,n_2}=\hbar(n_1+n_2+1)\phi_{n_1,n_2}\,,\qquad
 \hat{p}_\varphi\phi_{n_1,n_2}=\hbar(n_1-n_2)\phi_{n_1,n_2}\,.\label{Dphieig}
&
\end{eqnarray} 
From the first equation in (\ref{HyKonphi}),
 we see that  the action of the free particle Hamiltonian on $\phi_{n_1,n_2}$ 
 decreases both indexes $n_1$ and  $n_2$, and annihilates 
$\phi_{n_1,0}$ and $\phi_{0,n_2}$. 
It is clear then  that
functions (\ref{JordanStates}) 
are the zero energy Jordan states of $\hat{H}$. 
Using Eq. (\ref{Dphieig}), 
we find
the normalized wave functions  $\Psi_{n_1,n_2}=\mathcal{N}_{n_1,n_2}\hat{\mathfrak{S}}\phi_{n_1,n_2}$
  (where $\mathcal{N}_{n_1,n_2}$ is a numerical factor, see below) of $ \hat{H}_g $,
 \be
 \hat{H}_g \Psi_{n_1,n_2}=E_{n_1,n_2}\Psi_{n_1,n_2}\,,\qquad
 E_{n_1,n_2}=\hbar\omega (\ell_1 n_1+\ell_2 n_2+1)\,,
 \ee
 which simultaneously are eigenfunctions of the angular momentum operator,
 $\hat{p}_\varphi \Psi_{n_1,n_2}=\hbar(n_1-n_2)\Psi_{n_1,n_2}$.
Note that the spectrum of the system 
has degeneracies
if an only if $g$ is a rational number.
All energy levels are  positive
if $|g|\leq1$, while  
for $|g|>1$
 the energy levels can take negative values
and the 
spectrum is not bounded from below.
Moreover, for rational $g$ with  $|g|<1$  each energy level 
is finitely degenerate and the ground state with $n_1=n_2=0$ is
non-degenerate, while
for rational $g$ with $|g|\geq 1$ 
each energy level has infinite degeneracy.
In the limit case of the Landau phase $g=+1$ ($g=-1$),
one has $\ell_2=0$ ($\ell_1=0$),  and all the 
energy levels, including the lowest Landau level,
become infinitely degenerate.

 The explicit  action of the CBT operator $\hat{\mathfrak{S}}$ on functions $\phi_{n_1,n_2}$ is computed  
 by employing the inverse Weierstrass 
 transformation \cite{Bilo,InzPlyWipf1}
  \be
 e^{-\frac{1}{4}\frac{d^2}{d\eta^2}}\eta^{n}=2^{-n}H_{n}(\eta)\,,
 \ee
  where $H_{n}(\eta)$ are the Hermite polynomials. This
gives us the normalized  
eigenfunctions
\begin{eqnarray}\label{Psinn}
&
\Psi_{n_1,n_2}= 
\sqrt{\frac{m\omega}{\hbar \pi n_1!n_2!}} 
\,H_{n_1,n_2}\left(\sqrt{\frac{m\omega}{\hbar}}x_1,\sqrt{\frac{m\omega}{\hbar}}x_2\right)
e^{-\frac{m\omega}{2\hbar}(x_1^2+x_2^2)}\,,&\\&
\mathcal{N}_{n_1,n_2}=\left(\frac{2\hbar}{m\omega}\right)^{\frac{n_1+n_2}{2}} \sqrt{n_1!n_2! \pi}\,.
&
\end{eqnarray}
Here, the functions 
\begin{eqnarray}
&
H_{n_1,n_2}(\eta_1,\eta_2)=2^{n_1+n_2}\sum_{k=0}^{n_1}
\sum_{l=0}^{n_2}(i)^{n_1-n_2+l-k}H_{l+k}(\eta_1)H_{n_1+n_2 -l-k}(\eta_2)\,,&\\&
\int_{-\infty}^{\infty}\int_{-\infty}^{\infty}H_{n_1,n_2}(\eta_1,\eta_2)
H_{l_1,l_2}(\eta_1,\eta_2)e^{-(\eta_1^2+\eta_2^2)}d\eta_1d\eta_2=
\pi n_1! n_2!\delta_{n_1,l_1}\delta_{n_2,l_2}\,,
&
\end{eqnarray}
correspond to the generalized Hermite polynomials of two 
indexes \cite{Hermite}. 

Using the CBT relations 
\begin{eqnarray}
&
\hat{\mathfrak{S}}(\hat{p}_-)\hat{\mathfrak{S}}^{-1}=-i\sqrt{2m\hbar\omega}\,\hat{b}_1^-\,,\qquad
\hat{\mathfrak{S}}(\hat{p}_+)\hat{\mathfrak{S}}^{-1}=-i\sqrt{2m\hbar\omega}\,\hat{b}_2^-\,,
&\\&
\hat{\mathfrak{S}}(\hat{\xi}_{+})\hat{\mathfrak{S}}^{-1}=\sqrt{\frac{2m\hbar}{\omega}}\,\hat{b}_1^+\,,\qquad
\hat{\mathfrak{S}}(\hat{\xi}_-)\hat{\mathfrak{S}}^{-1}=\sqrt{\frac{2m\hbar}{\omega}}\,\hat{b}_2^+\,,
&
\end{eqnarray}
one finds 
the action of operators $\hat{b}_i^\pm$ on  eigenstates (\ref{Psinn}),
\begin{eqnarray}
&
\hat{b}_1^-\Psi_{n_1,n_2}=\sqrt{n_1}\Psi_{n_1-1,n_2}\,,\qquad
\hat{b}_2^-\Psi_{n_1,n_2}=\sqrt{n_2}\Psi_{n_1,n_2-1}\label{specgen1} &\\&
\hat{b}_1^+\Psi_{n_1,n_2}=\sqrt{n_1+1}\Psi_{n_1+1,n_2}\,,\qquad
\hat{b}_2^+\Psi_{n_1,n_2}=\sqrt{n_2+1}\Psi_{n_1,n_2+1}\,,\label{specgen2}&
\end{eqnarray}
wherefrom it is clear that 
they are spectrum generating operators. 

Let us study  the  spectral characteristics for the cases in which $g$ is a rational number.
We  do not consider the issue of the quantum algebra here
which inherits the  properties 
of the corresponding classical algebra, but
only identify  the integrals  associated with 
the spectral degenerations of the system.
\vskip0.25cm
\underline{Case $|g|<1$}.
Let us assume that
 $g=g^{s_1,s_2}_<=\frac{s_2-s_1}{s_2+s_1}$
  is an irreducible fraction  with some fixed 
non-negative integer  values of 
 $s_1$ and $s_2$. 
In this case the condition (\ref{condition 1}) is fulfilled, implying that 
$E_{n_1,n_2}=E_{n_1+js_1,n_2-js_2}$, where  $j$ is an integer number
such that $n_1+js_1\geq 0$, 
$n_2-js_2\geq 0$.
 One can construct the quantum  operators  
\begin{eqnarray}
\hat{\mathcal{L}}_{s_1,s_2}^+=(\hat{b}_1^+)^{s_1}(\hat{b}_2^-)^{s_2}\,,\qquad
\hat{\mathcal{L}}_{s_1,s_2}^-=(\hat{\mathcal{L}}_{s_1,s_2}^-)^\dagger \,,
\label{OperatorsA}
\end{eqnarray}
which are the direct quantum analogs of (\ref{higherorderA})
with $j_1=s_1$, $j_2=s_2$, $[\hat{H}_g,\hat{\mathcal{L}}_{s_1,s_2}^\pm]=0$.
These quantum integrals can be obtained by application of the 
conformal bridge transformation to the free particle higher 
order operators
\begin{equation}
\hat{S}_{s_1,s_2}^+=(\hat{\xi}_+)^{s_1}(\hat{p}_+)^{s_2}\,,\qquad
\hat{S}_{s_1,s_2}^-=(\hat{p}_-)^{s_1}(\hat{\xi}_-)^{s_2}\,,
\end{equation}
which commute with the operator $2i\hat{D}+g\hat{p}_\varphi$ with $g=g^{s_1,s_2}_<$.

Using relations (\ref{specgen1}) and (\ref{specgen2}) we get
\begin{eqnarray}
&
\hat{\mathcal{L}}_{s_1,s_2}^+\Psi_{n_1,n_2}=\sqrt{\frac{n_2!\Gamma(n_1+s_1+1)}{n_1!\Gamma(n_2-s_2+1)} }\Psi_{n_1+s_1,n_2-s_2}\,,&
\label{Lintdeg+}\\&
\hat{\mathcal{L}}_{s_1,s_2}^-\Psi_{n_1,n_2}=\sqrt{\frac{n_1!\Gamma(n_2+s_2+1)}{n_2!\Gamma(n_1-s_1+1)} }\Psi_{n_1-s_1,n_2+s_2}\,.
\label{Lintdeg-}&
\end{eqnarray} 
These equalities  imply that the operators $\hat{\mathcal{L}}_{s_1,s_2}^\pm$ allow us to obtain 
the complete  set of physical eigenstates which have the same energy 
but different angular momentum eigenvalues starting from some 
fixed eigenstate $\Psi_{n_1,n_2}$, i.e.,
 they correspond to integrals of motion 
 associated with hidden symmetries of the system
 which are  responsible for 
degeneracy of the spectrum. 
In  the cases $g=0$ and $g=\pm1$
we recover the symmetry operators of the isotropic harmonic oscillator and the Landau 
system in the symmetric gauge,
respectively.

\vskip0.25cm
\underline{Case $|g|>1$.} 
We suppose now that 
$g=g^{s_1,s_2}_>=\frac{s_2+s_1}{s_2-s_1}$
 is an irreducible fraction with  $s_1\not=s_2$. 
In this case the condition $s_1\ell_1+s_2\ell_2=0$ is fulfilled, implying that 
$E_{n_1,n_2}=E_{n_1+js_1,n_2+js_2}$,  
where now  $j$ is an integer number
such that $n_1+js_1\geq 0$, 
$n_2+js_2\geq 0$.
We can construct here  the operators
 \be
\hat{ \mathcal{J}}_{s_1,s_2}^+=(\hat{b}_1^+)^{s_1}(\hat{b}_2^+)^{s_2}\,,\qquad
 \hat{\mathcal{J}}_{s_1,s_2}^-=(\hat{\mathcal{J}}_{s_1,s_2}^-)^\dagger\,,
 \ee
which are the quantum analogs of the integrals  (\ref{higherorderB}) with $j_1=s_1$ and $j_2=s_2$,
$[\hat{H}_g,\hat{\mathcal{J}}_{s_1,s_2}^\pm]=0$.

These integrals are obtained by the application of the 
CBT to the free particle higher 
order operators
\begin{equation}
\hat{\Xi}_{s_1,s_2}^+=(\hat{\xi}_+)^{s_1}(\hat{\xi}_-)^{s_2}\,,\qquad
\hat{\Xi}_{s_1,s_2}^-=(\hat{p}_+)^{s_1}(\hat{p}_-)^{s_2}\,,
\end{equation}
which commute with the operator $2i\hat{D}+g\hat{p}_\varphi$ with 
$g=g^{s_1,s_2}_>$.
Their action on the eigenstates 
is given by
\begin{eqnarray}
&
\hat{\mathcal{J}}_{s_1,s_2}^-\Psi_{n_1,n_2}=\sqrt{\frac{n_1!n_2!}{\Gamma(n_1-s_1+1)
\Gamma(n_2-s_2+1)}}\Psi_{n_1-s_1,n_2-s_2}\,,\label{Jintdeg-}
&\\&
\hat{\mathcal{J}}_{s_1,s_2}^+\Psi_{n_1,n_2}=\sqrt{\frac{\Gamma(n_1+s_1+1)
\Gamma(n_2+s_2+1)}{n_1!n_2!}}\Psi_{n_1+s_1,n_2+s_2}\,.\label{Jintdeg+}
&
\end{eqnarray}
All the normalizable eigenfunctions  with the same energy can be obtained by repeated  application 
of these operators to some fixed state $\Psi_{n_1,n_2}$.
It is worth to note here the difference in the action of 
the integrals $\hat{\mathcal{L}}^\pm_{s_1,s_2}$ in the case  $g=\frac{s_2-s_1}{s_2+s_1}$, 
and integrals $\hat{\mathcal{J}}^\pm_{s_1,s_2}$ for  $g=\frac{s_2+s_1}{s_2-s_1}$.
In the first case, 
after repeated application  of the corresponding integral operator with index 
plus or minus, at some step we obtain zero 
due to appearance of a  Gamma function pole in denominator of 
coefficients in (\ref{Lintdeg+}) and (\ref{Lintdeg-}). 
A similar situation we have only for the action of 
the operator
$\hat{\mathcal{J}}^-_{s_1,s_2}$ in the case of $g=\frac{s_2+s_1}{s_2-s_1}$,
but the repeated application of the operator $\hat{\mathcal{J}}^+_{s_1,s_2}$
will never produce zero, see the coefficient in Eq. (\ref{Jintdeg+}).
This difference is just another reflection
of the finite and infinite degeneracy of energy levels 
in the two indicated cases.

In conclusion of this section, we show how the CBT 
can be used  to construct  coherent states for  the system
$\hat{H}_g$. 
For this  
we consider the exponential function $e^{\frac{1}{\sqrt{2}}(\alpha z+ \beta z^*)}$,
which, in dependence on the  values of the parameters 
$\alpha,\beta\in \C$  is either the plane wave eigenfunction, or formal, non-physical 
eigenfunction of the free particle Hamiltonian operator $\hat{H}$.
 The application  
of the CBT operator yields
\begin{eqnarray}
\begin{array}{lcl}
\mathfrak{S}e^{\frac{1}{\sqrt{2}}(\alpha z+ \beta z^*)}&=&
\sqrt{2}e^{-\frac{m\omega}{2}zz^*+\alpha z+\beta z^* -\frac{\hbar}{m\omega}\alpha\beta}\\
&=&\sum_{n=0}^{\infty}\sum_{k=0}^{n}\left(\frac{\hbar}{m\omega}\right)^{\frac{n}{2}}
\frac{\alpha^{n-k}\beta^k}{\sqrt{k!(n-k)!}}\alpha^{n-k}\beta^k\Psi_{n-k,k}\\
&=& \mathcal{C}_{\alpha,\beta}\Phi(x_1,x_2,\alpha,\beta)\,,
\end{array}
\end{eqnarray}
where $\mathcal{C}_{\alpha,\beta}=\sqrt{\pi}e^{-\frac{\hbar}{2m\omega}(|\alpha|^2+|\beta|^2)}$,
 and 
$\Phi(x_1,x_2,\alpha,\beta)$ is a normalized function. 
The  expansion of  $\Phi(x_1,x_2,\alpha,\beta)$ over  the orthonormal eigenstates of the system allows us  
to see how these states transform under time translations and rotations,
\begin{eqnarray}
&
e^{-\frac{it}{\hbar}\hat{H}_g}\Phi(x,y,\alpha,\beta)=\Phi(x,y,\alpha e^{-i\omega \ell_1 t},\beta e^{-i\omega \ell_2 t})\,,&\\&
e^{\frac{i\gamma}{\hbar}\hat{p}_\varphi }\Phi(x,y,\alpha,\beta)=\Phi(x,y,\alpha e^{i\gamma},\beta e^{-i\gamma})\,.&
\end{eqnarray}
On the other hand,  
 the introduced exponential is an eigenstate of the operators $\hat{p}_\pm$, 
 and then one has  
\begin{eqnarray}
&
\hat{b}_1^-\Phi(x,y,\alpha,\beta)=\sqrt{\frac{\hbar}{m\omega }}\alpha \Phi(x,y,\alpha,\beta)\,,\quad
\hat{b}_2^-\Phi(x,y,\alpha,\beta)=\sqrt{\frac{\hbar}{m\omega }}\beta \Phi(x,y,\alpha,\beta)\,.
&
\end{eqnarray}
 As these functions are the eigenstates of the lowering operators $\hat{b}_i^-$, and 
 they maintain  their form  without dispersion while time evolves,  we conclude that 
 $\Phi(x,y,\alpha,\beta)$ are coherent states 
 for the system $\hat{H}_g$.  
 
\section{Generation of AHO by anisotropic CBT}
\label{SecAHOINCBT}

In  
the previous section 
 we have discussed the ERIHO 
generated from
the two-dimensional free particle by the 
rotationally invariant CBT. 
In this section we explore the possibility of connecting the free particle 
and the AHO which does not 
have rotational invariance. For this, we employ an 
 anisotropic 
 CBT. 
 
Consider the following generator of 
the  CBT 
(no summation over 
 the repeated index) 
\be\label{Sanis}
\hat{\mathfrak{S}}_{\omega_1,\omega_2}=\hat{\mathfrak{S}}_{\omega_1}\hat{\mathfrak{S}}_{\omega_2}\,,\qquad 
\hat{\mathfrak{S}}_{\omega_i}=e^{-\frac{\omega_i}{\hbar} \hat{K}_i} 
e^{\frac{\hat{H}_i}{2\hbar \omega_i}}
e^{\frac{i}{\hbar}\ln(2)\hat{D}_i}\,.
\ee
In the case $\omega_1\neq \omega_2$, this generator 
is anisotropic (rotationally non-invariant) operator.
Instead of the linear combination of the free particle 
symmetry generators $\omega (2i\hat{D}+g\hat{p}_\varphi)$,
we apply the anisotropic CBT to the two
operators 
\be\label{2iDani}
2i\hat{\mathcal{D}}_\pm:=
2i(\omega_1\hat{D}_1\pm \omega_2\hat{D}_2)\,.
\ee
By using the decomposition (\ref{TotalSo(2,1)}) and the second  relation in (\ref{GenConBr}),
we obtain \footnote{As before, one could introduce the notation 
$\ell_1=1+g$, $\ell_2=1-g$, 
$g\in\R$, $\omega_i=\ell_i\omega$, and  instead of (\ref{2iDani}),
apply the CBT (\ref{Sanis}) to the 
operator $2i\omega\hat{\mathcal{D}}_g:=
2i\omega(\ell_1D_1+\ell_2D_2)$. This, however, 
will not change 
the final results, see
 below.}
 \begin{eqnarray}
&\label{InhomoGenConBr}
\hat{\mathfrak{S}}_{\omega_1,\omega_2}
(2i\hat{\mathcal{D}}_\pm)
\hat{\mathfrak{S}}_{\omega_1,\omega_2}^{-1}=\sum_{i=1}^{2}(\pm 1)^{i-1}\left(
-\frac{\hbar^2}{2m}\frac{\partial^2}{\partial x_i^2}+\frac{m\omega_i^2}{2}x_i^2
\right):=\hat{H}_{\omega_1,\omega_2}^{(\pm)}\,.&
\end{eqnarray}
When we choose the positive  sign, 
the operator $\hat{H}_{\omega_1,\omega_2}^{(+)}$
 corresponds to the  Hamiltonian operator  of the usual, Euclidean 
 AHO system. When the negative sign is selected, 
  $\hat{H}_{\omega_1,\omega_2}^{(-)}$ corresponds 
 to the  Minkowskian AHO.

Acting on the momenta operators and the Galilean boost generators of the free particle, 
the anisotropic  CBT produces
\begin{eqnarray}
\label{mappingSG}
&
\hat{\mathfrak{S}}_{\omega_1,\omega_2}(\hat{p}_i)
\hat{\mathfrak{S}}_{\omega_1,\omega_2}^{-1}=
-i\sqrt{m\hbar\omega_i}\,\hat{a}_{\omega_i}^-\,,\qquad
\hat{\mathfrak{S}}_{\omega_1,\omega_2}(\hat{\xi}_{i})
\hat{\mathfrak{S}}_{\omega_1,\omega_2}^{-1}=
\sqrt{\frac{m\hbar}{\omega_i}}\,\hat{a}_{\omega_i}^+\,,
&
\end{eqnarray} 
where
 \begin{eqnarray}
 &
 \label{aomega}
\hat{a}_{\omega_i}^\pm=\sqrt{\frac{m\omega_i}{2\hbar}}\left(x_i \mp \frac{\hbar}{m\omega_i}
\frac{\partial}{\partial x_i}\right)
&
 \end{eqnarray}
are the first-order ladder operators for each direction.
From here, we find that the angular momentum is not invariant 
under the anisotropic CBT,
\begin{eqnarray}
&
\hat{\mathfrak{S}}_{\omega_1,\omega_2}
(\hat{p}_\varphi)\hat{\mathfrak{S}}_{\omega_1,\omega_2}^{-1}=-i\hbar\left(\sqrt{\frac{\omega_2}{\omega_1}}
\hat{a}^+_{\omega_1}\hat{a}^-_{\omega_2}-\sqrt{\frac{\omega_1}{\omega_2}}
\hat{a}^+_{\omega_2}\hat{a}^-_{\omega_1}\right)\,.
&
\end{eqnarray}
Instead of the angular momentum operator in the case of $\hat{H}_g$,
the AHO systems 
\begin{eqnarray}
&
\hat{H}_{\omega_1,\omega_2}^{(\pm)}=\hbar
\left(\omega_1\hat{a}^+_{\omega_1}\hat{a}^-_{\omega_1}
\pm\omega_2 \hat{a}^+_{\omega_2}\hat{a}^-_{\omega_2}
+\frac{1}{2}(\omega_1\pm
\omega_2)\right)
&
\end{eqnarray}
are characterized here by the obvious symmetry generators
\begin{eqnarray}\label{L+-omom}&
\hat{L}_{\omega_1,\omega_2}^{(\pm)}=\hbar
\left(\omega_1\hat{a}^+_{\omega_1}\hat{a}^-_{\omega_1}
\mp\omega_2 \hat{a}^+_{\omega_2}\hat{a}^-_{\omega_2}\right)\,.&
\end{eqnarray}
Therefore,  in the case  $\omega_1=\omega_2=\omega$ ($\hat{a}_{\omega_i}^\pm\rightarrow\hat{a}_i^\pm$), the Hamiltonian of 
the isotropic, in the sense of $(1+1)$-dimensional Lorentzian metric,   Minkowskian 
oscillator
\be\label{Minso(1,1)}
\hat{H}^{(-)}_{\omega,\omega}=\hbar\omega(\hat{a}^+_1\hat{a}^-_1
-\hat{a}^+_2\hat{a}^-_2)
 \ee
 is invariant under the  $\mathfrak{so}(1,1)$  transformations generated 
 by the operator
 \be\label{L11}
 \hat{L}_{1,1}=\hat{x}_1\hat{p}_2+\hat{x}_2\hat{p}_1=
 i\hbar(\hat{\mathscr{J}}_+-\hat{\mathscr{J}}_-)\,,\qquad
 \hat{\mathscr{J}}_\pm=\hat{a}^\pm_1\hat{a}^\pm_2\,.
 \ee
The true integrals $\hat{\mathscr{J}}_\pm$, together with 
\footnote{ In the context of the symmetry transmutation, the planar isotropic harmonic oscillator
Hamiltonian  appears here as the  integral of the system we are dealing with. 
It  is invariant under  $\mathfrak{su}(2)$ transformations  
generated by  (\ref{u1osc}) taken with $g=0$, which for the system 
(\ref{Minso(1,1)}) correspond to the explicitly  depending on time,  
dynamical integrals.}
$\hat{\mathscr{J}}_{0}=\frac{1}{2\omega\hbar}L^{(-)}_{\omega,\omega}
=\frac{1}{2\omega\hbar}(\hat{H}_{\text{osc}}-\hbar\omega)$  
produce the $\mathfrak{sl}(2,\R)$ symmetry algebra 
of the system (\ref {Minso(1,1)}). 
As $\hat{\mathscr{J}}_{0}$
and the combination 
$\hbar(\hat{\mathscr{J}}_-+\hat{\mathscr{J}}_+)=
\frac{\hat{p}_1\hat{p}_2}{2m}-m\omega\hat{x}_1\hat{x}_2
$ 
are the second
order in momenta integrals of motion, they correspond to
the  hidden symmetry 
operators \cite{Cariglia}. 

The obvious choice for the well defined in $\R^2$ 
eigenfunctions that obey  the eigenvalue equation 
$
2i\hat{\mathcal{D}}_\pm \phi_{\lambda}=\lambda\phi_{\lambda}
$
corresponds to 
\be
\phi_{n_1,n_2}=x_1^{n_1}x_2^{n_2}\,
\ee
with non-negative integer values of $n_1$ and $n_2$.
These eigenfunctions  satisfy the relations 
\begin{eqnarray}
&\label{pyxionphiani}
\hat{p}_1\phi_{n_1,n_2}=-i\hbar n_1\phi_{n_1-1,n_2}\,,\qquad
\hat{p}_2\phi_{n_1,n_2}=-i\hbar n_2\phi_{n_1,n_2-1}\,,&\\&
\hat{\xi_1}\phi_{n_1,n_2}=m\phi_{n_1+1,n_2}\,,\qquad
\hat{\xi_2}\phi_{n_1,n_2}=m\phi_{n_1,n_2+1}\,.\label{pyxionphiani2}
\end{eqnarray}
They  are the zero energy Jordan states (generally, of different orders
in the case of $n_1\neq n_2$) of the one-dimensional
Hamiltonian operators $\hat{H}_i$, 
and the action of $\hat{\mathfrak{S}}_{\omega_1,\omega_2}$ on them yields 
\begin{eqnarray}
&
\hat{\mathfrak{S}}_{\omega_1,\omega_2}\phi_{n_1,n_2}(x_1,x_2)=
(\frac{\omega_1}{\omega_2})^{-\frac{1}{4}}(\frac{\hbar}{m})^{\frac{n_1+n_2+1}{2}}\sqrt{2n_1!n_2!\pi}\,
\psi_{n_1,n_2}(x_1,x_2)\,,
&\\&
\psi_{n_1,n_2}(x_1,x_2)=\psi_{n_1}(x_1)\psi_{n_2}(x_2)\,,&\\&
\psi_{n_i}(x_i)=\frac{1}{\sqrt{2^{n_i}n_i!}}(\frac{m\omega_i}{\pi \hbar})^{\frac{1}{4}}
H_{n_i}(\sqrt{\frac{m\omega_i}{\hbar}}x_i)e^{-\frac{m\omega_i}{2\hbar}x_i^2}\,,
&
\end{eqnarray}
where $H_{n_i}$ are the Hermite polynomials.
Meanwhile, the formal eigenvalue equation 
\begin{eqnarray}
&
2i\hat{\mathcal{D}}_\pm\phi_{n_1,n_2}=
\hbar\left(\omega_1 n_1\pm \omega_2 n_2+\frac{\omega_1\pm\omega_2}{2}\right)\phi_{n_1,n_2}&
\end{eqnarray}
implies that 
\begin{eqnarray}&
\hat{H}_{\omega_1,\omega_2}^{(\pm)}\psi_{n_1,n_2}=\hbar\left(\omega_1 n_1\pm \omega_2 n_2+
\frac{\omega_1\pm
\omega_2}{2}\right)\psi_{n_1,n_2}\,. &
\end{eqnarray}
From here it follows that  the energy values  of the Euclidean AHO  $\hat{H}_{\omega_1,\omega_2}^{(+)}$
are positive, while  in the case of Minkowskian 
AHO  described by $\hat{H}_{\omega_1,\omega_2}^{(-)}$
the spectrum is not bounded from below.

From  relations (\ref{mappingSG}), (\ref{pyxionphiani}) and (\ref{pyxionphiani2}),
one concludes  that (\ref{aomega}) 
are the spectrum generating 
operators that satisfy relations  
\begin{eqnarray}
&
\hat{a}^-_{\omega_1}\psi_{n_1,n_2}=\sqrt{n_1}\psi_{n_1-1,n_2}\,,\qquad
\hat{a}^-_{\omega_2}\psi_{n_1,n_2}=\sqrt{n_2}\psi_{n_1,n_2-1}\,,\label{specgen1s} &\\&
\hat{a}^+_{\omega_1}\psi_{n_1,n_2}=\sqrt{n_1+1}\psi_{n_1+1,n_2}\,,\qquad
\hat{a}^+_{\omega_2}\psi_{n_1,n_2}=\sqrt{n_2+1}\psi_{n_1,n_2+1}\,.\label{specgen2s}&
\end{eqnarray}

On the other hand,  it is well known
that besides the integrals (\ref {L+-omom}),
both  systems $\hat{H}_{\omega_1,\omega_2}^{(\pm)}$
have additional, higher order true integrals of motion when  frequencies 
are commensurable, $\omega_1/\omega_2=l_2/l_1$ \cite{BHT}.
In the case of $\hat{H}_{\omega_1,\omega_2}^{(+)}$, these integrals can be obtained by 
applying the anisotropic  CBT to the 
higher order operators
\begin{equation}
\hat{\mathcal{S}}_{l_1,l_2}=(\hat{\xi}_1)^{l_1}(\hat{p}_2)^{l_2}\,,\qquad
\hat{\mathcal{S}}_{l_2,l_1}=(\hat{p}_1)^{l_1}(\hat{\xi}_2)^{l_2}\,,
\end{equation}
which commute with $\hat{\mathcal{D}}_+$ in this case. One has 
\begin{eqnarray}
&
\hat{\mathfrak{S}}_{\omega_1,\omega_2}(\hat{\mathcal{S}}_{l_1,l_2})\hat{\mathfrak{S}}_{\omega_1,\omega_2}^{-1}\propto 
\hat{\mathscr{L}}_{l_1,l_2}^{+}\,.
\qquad
\hat{\mathfrak{S}}_{\omega_1,\omega_2}(\hat{\mathcal{S}}_{l_2,l_1})\hat{\mathfrak{S}}_{\omega_1,\omega_2}^{-1}\propto 
\hat{\mathscr{L}}_{l_1,l_2}^-\,,
&\\&
\hat{\mathscr{L}}_{l_1,l_2}^+=(\hat{a}^+_{\omega_1})^{l_1}(\hat{a}^-_{\omega_2})^{l_2}\,,\qquad
\hat{\mathscr{L}}_{l_1,l_2}^-=(\hat{\mathscr{L}}_{l_1,l_2}^+)^\dagger=\hat{\mathscr{L}}_{l_2,l_1}^+\,.
&
\end{eqnarray}
The explicit action of these hidden 
symmetry operators is given by   
\begin{eqnarray}
&\label{Ll1l2-}
\hat{\mathscr{L}}_{l_1,l_2}^-\psi_{n_1,n_2}=
\sqrt{\frac{n_1!\Gamma(n_2+l_2+1)}{n_2!\Gamma(n_1-l_1+1)} }\psi_{n_1-l_1,n_2+l_2}\,,&\\&
\hat{\mathscr{L}}_{l_1,l_2}^+\psi_{n_1,n_2}=
\sqrt{\frac{n_2!\Gamma(n_1+l_1+1)}{n_1!\Gamma(n_2-l_2+1)} }\psi_{n_1+l_1,n_2-l_2}\,.
&\label{Ll1l2+}
\end{eqnarray}
In the special case in which $\omega_1=\omega_2$, implying $l_1=l_2=1$, 
integral operators $\hat{\mathscr{L}}_{1,1}^\pm$ and 
$\hat{\mathscr{L}}_2^{(+)}=
\frac{1}{2}\hat{p}_\varphi$,
to which the operator $\frac{1}{2\omega}\hat{L}_{\omega,\omega}^{(+)}$
 from (\ref{L+-omom}) is reduced, 
 generate the $\mathfrak{su}(2)$ algebra. 
 For $l_1\neq l_2$, the integrals $\hat{L}_{\omega_1,\omega_2}^{(+)}$
 and $\hat{\mathscr{L}}_{l_1,l_2}^\pm$ together with Hamiltonian 
 $\hat{H}_{\omega_1,\omega_2}^{(+)}$
  generate a non-linear deformation of $\mathfrak{u}(2)$.
 Classical analogs of $\hat{\mathscr{L}}_{l_1,l_2}^\pm$
  correspond to hidden symmetries 
because they generate  the transformations that    
mix coordinates and momenta  in phase space.  
 
In the case of $\hat{H}_{\omega_1,\omega_2}^{(-)}$, 
additional true integrals 
are obtained from the operators 
\begin{eqnarray}
\hat{\mathcal{P}}_{l_1,l_2}=(\hat{p}_1)^{l_1} (\hat{p}_2)^{l_2}\,,\qquad
\hat{\Xi}_{l_1,l_2}=(\hat{\xi}_1)^{l_1} (\hat{\xi}_2)^{l_2}\,,
\end{eqnarray}
which commute with
 $\hat{\mathcal{D}}_-$. They are transformed by the anisotropic CBT into 
\begin{eqnarray}
&
\hat{\mathfrak{S}}_{\omega_1,\omega_2}(\hat{\Xi}_{l_1,l_2})\hat{\mathfrak{S}}_{\omega_1,\omega_2}^{-1}\propto 
\hat{\mathscr{J}}_{l_1,l_2}^{+}\,\qquad
\hat{\mathfrak{S}}_{\omega_1,\omega_2}(\hat{\mathcal{P}}_{l_1,l_2})\hat{\mathfrak{S}}_{\omega_1,\omega_2}^{-1}\propto 
\hat{\mathscr{J}}_{l_1,l_2}^-\,,\qquad
 &\\&
\hat{\mathscr{J}}_{l_1,l_2}^+=(\hat{a}^+_{\omega_1})^{l_1}(\hat{a}^+_{\omega_2})^{l_2}\,,\qquad
\hat{\mathscr{J}}_{l_1,l_2}^-=(\hat{\mathscr{J}}_{l_1,l_2}^+)^\dagger\,.
&
\end{eqnarray}
They act 
 on the eigenstates $\psi_{n_1,n_2}(x_1,x_2)$ as follows,
\begin{eqnarray}
&
\label{Jl1l2-}
\mathscr{J}_{l_1,l_2}^-\psi_{n_1,n_2}=\sqrt{\frac{n_1!n_2!}{\Gamma(n_1-l_1+1)\Gamma(n_2-l_2+1)}}\psi_{n_1-1,n_2-1}\,,
&\\&
\mathscr{J}_{l_1,l_2}^+\psi_{n_1,n_2}=\sqrt{\frac{\Gamma(n_1+l_1+1)\Gamma(n_2+l_2+1)}{n_1!n_2!}}\psi_{n_1+1,n_2+1}\,.
&
\label{Jl1l2+}
\end{eqnarray}
In the case $l_1=l_2=1$ ($\omega_1=\omega_2=\omega$), the system 
$\hat{H}_{\omega,\omega}^{(-)}$, as we already noted, 
corresponds to the  $\mathfrak{so}(1,1)$-invariant Minkowskian oscillator
(\ref{Minso(1,1)}), for which  
the not depending explicitly on time  integrals (\ref{L11}), 
$\mathscr{J}_{1,1}^\pm=\mathscr{J}_\pm$, 
$\hat{\mathscr{J}}_{0}=\frac{1}{2\omega\hbar}(\hat{H}_{\text{osc}}-\hbar\omega)$ and 
$\hat{H}_{\omega,\omega}^{(-)}$
generate the $\mathfrak{gl(}2,\R)$ symmetry.
Note that analogously to  the ERIHO, here
the bounded and unbounded from below character of the spectra in the cases  
of Euclidean and Minkowskian AHO systems
is encoded   in the structure of coefficients in 
Eqs. (\ref{Ll1l2-}), (\ref{Ll1l2+}) and  (\ref{Jl1l2-}),  (\ref{Jl1l2+}), 
respectively.

By applying the anisotropic CBT to the physical,  or to non-physical
eigenstates $e^{\frac{1}{\sqrt{2}}(\alpha_1 x_1+\alpha_2 x_2)}$,
$\alpha_1,\, \alpha_2\in \C$, of the free particle Hamiltonian, we obtain
\begin{eqnarray}
&
\begin{array}{lcl}
\Phi(x_1,x_2,\alpha_1,\alpha_2)&:=&\hat{\mathfrak{S}}_{\omega_1,\omega_2}e^{\frac{1}{\sqrt{2}}(\alpha_1 x_1+\alpha_2 x_2)}\\
&=&\sqrt{2}e^{-\frac{\hbar}{4m}(\frac{\alpha_1^2}{\omega_1}+\frac{\alpha_2^2}{\omega_2})-
\frac{m}{2\hbar}(\omega_1 x_1^2+\omega_2 x_2^2)+\alpha_1 x_1+\alpha_2 x_2} \,.
\end{array}&
\end{eqnarray}
These functions correspond to 
coherent states for both systems $\hat{H}_{\omega_1,\omega_2}^{(\pm)}$ as they  
 satisfy the relations
\begin{eqnarray}
&
\hat{a}^-_{\omega_i}\Phi(x,y,\alpha_1,\alpha_2)=\sqrt{\frac{\hbar}{2m\omega_i }}\alpha_i \Phi(x,y,\alpha,\beta)\,.
&
\end{eqnarray}

\begin{figure}[H]
\begin{center}
\begin{subfigure}[c]{0.28\linewidth}
\includegraphics[scale=0.5]{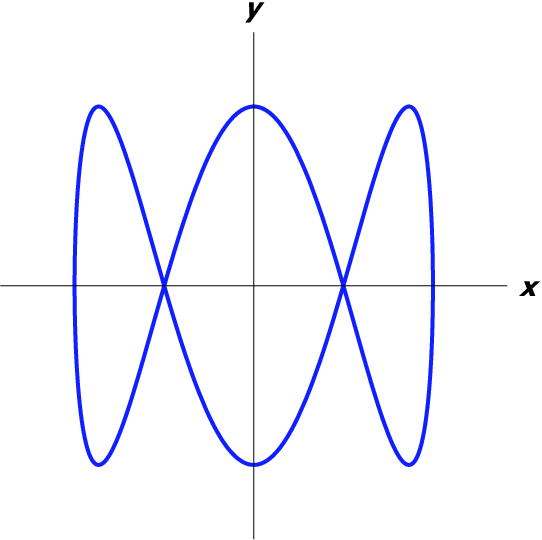}
\caption{\small{$\omega_1/\omega_2=1/3$}}
\end{subfigure}
\begin{subfigure}[c]{0.28\linewidth}
\includegraphics[scale=0.5]{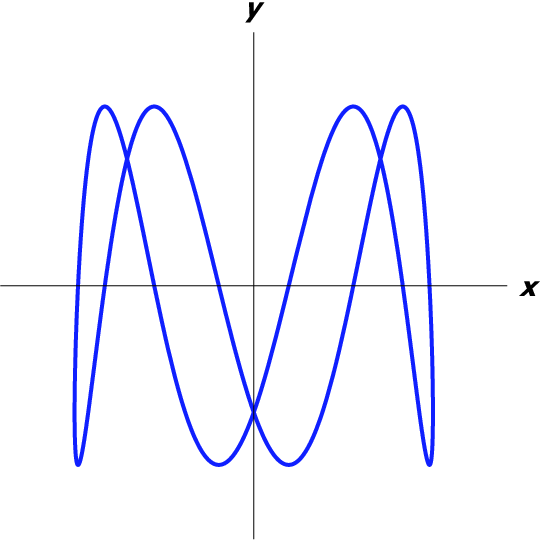}
\caption{\small{$\omega_1/\omega_2=1/4$}}
\end{subfigure}
\begin{subfigure}[c]{0.28\linewidth}
\includegraphics[scale=0.5]{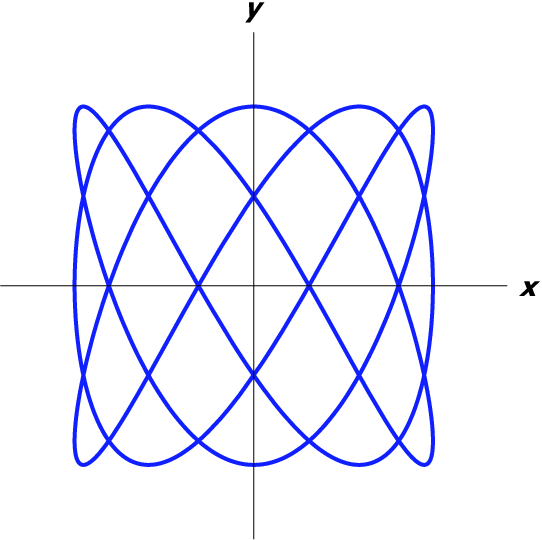}
\caption{\small{$\omega_1/\omega_2=3/5$}}
\end{subfigure}
\end{center}
\caption{\small{Trajectories of the anisotropic harmonic oscillator
with commensurable frequencies.}}
\label{Fig3}
\end{figure}

When comparing these results with the quantum analysis of the 
ERIHO systems, we observe that both models are really similar. 
Their spectra in Euclidean and Minkowskian cases have
 similar characteristics,  
while the corresponding spectrum-generating operators and the integral operators 
in both classes of models act in a similar  way on their respective eigenstates. 
At the same time we note that the difference in their properties with respect
to the planar rotations reveals itself 
in classical 
dynamics. This  can be observed by comparing the  form of 
trajectories
in the ERIHO systems with those 
in the systems discussed in the present section, for which 
trajectories are 
described by equations
\be\label{Lis}
x_1(t)=A_1\cos(\omega_1 t)+B_1\sin(\omega_1 t)\,,\qquad
x_2(t)=A_2\cos(\omega_2 t)+B_2\sin(\omega_2 t)\,.
\ee
In the case of commensurable frequencies these trajectories 
 also are closed but they  are represented by Lissajous curves,
some examples of which  are shown 
 in  Fig \ref{Fig3}. 
Essential difference also consists in the fact that 
equations (\ref{Lis}) describing the trajectories
have exactly the same form 
for  the Euclidean, $H^{(+)}_{\omega_1,\omega_2}$,
and Minkowskian,
$H^{(-)}_{\omega_1,\omega_2}$,
 AHOs with the same values
of frequencies $\omega_1$ and $\omega_2$
(though the same trajectories correspond to different energies
in the two indicated cases), 
while the form of dynamics and corresponding trajectories 
Euclidean ($|g|<1$) 
and Minkowskian ($|g|>1$)
ERIHO systems is different. 

Nevertheless, 
the similarities at the quantum level indicate 
 that the two classes of the harmonic oscillator systems should be related somehow, and in the 
 next section we describe their  relationship.

\section{Relationships of  the ERIHO}
\label{SecOtherS}

With the help of a unitary transformation, which corresponds to a rotation in the 
three-dimensional 
``ambient space"
 of the 
$ \mathfrak{su} (2) $ algebra, in this section we show that
the ERIHO and AHO  can be related to each other. 
We also relate the ERIHO 
with $g^2<1$ and $g^2>1$
with the Landau problem in the presence of the additional 
attractive  and repulsive  
harmonic potentials. 

\subsection{Relationship of 
the ERIHO and AHO systems}

Let us consider again the ERIHO Hamiltonian  
 \be
 \label{QmrotHg}
 \hat{H}_g= \hbar\omega(\ell_1\hat{b}_1^+\hat{b}_1^-  + \ell_2\hat{b}_2^+\hat{b}_2^- +1)\,,\qquad
 \ell_1=1+g\,,\qquad \ell_2=1-g\,,
 \ee
 and  introduce the 
 unitary operator \cite{InzPly7}
\begin{eqnarray}
\label{UnitaryRepChange}
&
\hat{U}=\exp(-i\frac{2\pi}{3}\frac{1}{\sqrt{3}}(\hat{\mathcal{L}}_1+\hat{\mathcal{L}}_2+\hat{\mathcal{L}}_3) )\,,&\\&
\mathcal{L}_1=\frac{1}{2}(\mathcal{L}_-
+
\mathcal{L}_+)=
\frac{1}{2}(\hat{a}_1^+\hat{a}_2^-+\hat{a}_2^+\hat{a}_1^-)\,,\quad
\mathcal{L}_3=
\frac{i}{2}(\mathcal{L}_-
-
\mathcal{L}_+)=
\frac{1}{2}(\hat{a}_1^+\hat{a}_1^-- \hat{a}_2^+\hat{a}_2^-)\,,
&
\end{eqnarray}
see Eqs. (\ref{u1osc})  and (\ref{barel}).
It  produces the unitary transformation 
 \begin{eqnarray}
&\label{abuni}
 \hat{U}\hat{a}_j^\pm\hat{U}^\dagger= e^{\pm i\frac{\pi}{4}}\hat{b}_j^\pm\,,&\\
&\hat{U}\hat{\mathcal{L}}_1 \hat{U}^{\dagger}=\hat{\mathcal{L}}_3\,,\quad
\hat{U}\hat{\mathcal{L}}_2 \hat{U}^{\dagger}=\hat{\mathcal{L}}_1\,,\quad
\hat{U}\hat{\mathcal{L}}_3 \hat{U}^{\dagger}=\hat{\mathcal{L}}_2\,.&\label{su2trans}
 \end{eqnarray}
 Using these relations, we find  that 
 \begin{eqnarray}
\label{Haniso}
\hat{H}_g=\hat{U}(\hat{H}_g^{\text{n-i}})\hat{U}^{\dagger}\,,
\qquad \text{where}\quad
\hat{H}_g^{\text{n-i}}=\hbar\omega(\ell_1\hat{a}_1^+\hat{a}_1^-  + \ell_2\hat{a}_2^+\hat{a}_2^- +1)\,.
\end{eqnarray}
Unlike  $\hat{H}_g$, 
Hamiltonian operator $\hat{H}_g^{\text{n-i}}$  is not invariant 
under the $\mathfrak{so}(2)$ rotations.
 In terms of the coordinates and momenta operators, 
 $\hat{H}_g^{\text{n-i}}$ is presented 
 as 
 \be
\hat{H}_g^{\text{n-i}}= \frac{1}{2m}\left(\ell_1\hat{p}_{1}^2+\ell_2
\hat{p}^2_{2}
\right)+
\frac{m\omega^2}{2}(\ell_1x_1^2+\ell_2 x_2^2)\,.
\ee 
After
 the  quantum canonical transformation 
of anisotropic rescaling,
\be\label{canxpl}
x_i\rightarrow x_i'=\sqrt{|\ell_i|} \,x_i\,,\qquad 
\hat{p}_i\rightarrow \hat{p}_i'=\hat{p}_i/\sqrt{|\ell_i|}\,,
\ee
we recognize  
\begin{eqnarray}
&\label{Hg''}
\hat{H}_g^{\text{n-i}'}=\text{sign }(\ell_1)\left(\frac{\hat{p'}_{1}^2}{2m}+\frac{m\Omega_1^2}{2}x_1'^2 \right)+
\text{sign}(\ell_2)\left(\frac{\hat{p'}_{1}^2}{2m}+\frac{m\Omega_2^2}{2}x_2'^2 \right)\,,\qquad
\Omega_i=|\ell_i |\omega\,. 
&
\end{eqnarray}
When $|g|<1$, the operator $\hat{H}_g^{\text{n-i}'}$ can be interpreted 
as the Hamiltonian of  a  one-parametric family of 
Euclidean  AHO systems. 
On the other hand, when
$|g|>1$ we have a one-parametric family of 
  Minkowskian 
AHO models.
 
By applying   unitary transformation inverse to 
(\ref{canxpl}),
 and then the transformation  inverse to (\ref{abuni}), (\ref{su2trans}),
to  the true and dynamical  integrals of the system $\hat{H}_g$, one gets, in particular,   
\begin{eqnarray}
&
\hat{b}_i^\pm\rightarrow \hat{a}_{\Omega_i}^\pm\,,\qquad
\hat{\mathcal{L}}_{s_1,s_2}^{\pm} \rightarrow 
\hat{\mathscr{L}}_{s_1,s_2}^{\pm}\,,\qquad
\hat{\mathcal{J}}_{s_1,s_2}^{\pm} \rightarrow 
\hat{\mathscr{J}}_{s_1,s_2}^{\pm}\,,&\\&
\hat{\mathcal{J}}_\pm\rightarrow\hat{\mathscr{J}}_\pm^{\Omega_1,\Omega_2}=
\hat{a}_{\Omega_1}^\pm\hat{a}^\pm_{\Omega_2}\,,\qquad
\hat{\mathcal{L}}_\pm\rightarrow\hat{\mathscr{L}}_\pm^{\Omega_1,\Omega_2}=
\hat{a}^\pm_{\Omega_1}\hat{a}^\mp_{\Omega_2}\,,&\\&
\hat{\mathcal{J}}_0-\frac{1}{2}\rightarrow\hat{\mathscr{J}}_0^{\Omega_1,\Omega_2}=
\frac{1}{2}(\hat{a}^+_{\Omega_1}\hat{a}^-_{\Omega_1}+
\hat{a}^+_{\Omega_2}\hat{a}^-_{\Omega_2})\,,&\\&
\hat{\mathcal{L}}_2\rightarrow\hat{\mathscr{L}}_2^{\Omega_1,\Omega_2}=
\frac{1}{2}(\hat{a}^+_{\Omega_1}\hat{a}^-_{\Omega_1}-\hat{a}^+_{\Omega_2}\hat{a}^-_{\Omega_2})\,.
&
\end{eqnarray}
So, we find that both, the ERIHO and AHO systems 
are unitary equivalent being 
related by the described composition of the  two unitary
transformations. The peculiarity of this relation is that 
(\ref{abuni}) and (\ref{su2trans})
corresponds to a particular $\mathfrak{su}(2)$ rotation 
 in 
a fictitious  
 three-dimensional 
space corresponding to the index of the
$\mathfrak{su}(2)$ generators, 
and the rotational invariance in the $\R^2$
configurational space is broken by the anisotropy 
of the rescaling transformation (\ref{canxpl}).
Notice that   unitary transformation (\ref{canxpl}) 
generated by the operator  $\exp(i(\hat{D}_1 \ln |\ell_1|+\hat{D}_2 \ln |\ell_2|))$ is
a Bogolyubov transformation \cite{Bogol} corresponding 
to a hyperbolic $\mathfrak{so}(1,1)\oplus \mathfrak{so}(1,1)$ rotation in terms of
the operators $(\hat{a}^+_1,\,\hat{a}^-_1)$, $(\hat{a}^+_2,\,\hat{a}^-_2)$.
In order to relate the  
isotropic Minkowskian case of the ERIHO
with its $\mathfrak{so}(1,1)$
analog (\ref{Minso(1,1)}), it is necessary
to make a change $\omega\rightarrow \omega/|g|$ in (\ref{Hg''}),
and then take a limit $|g|\rightarrow \infty$.

According to Eq. (\ref{abuni}),
the classical analog of the unitary transformation generated by 
(\ref{UnitaryRepChange}) mixes coordinates and momenta
variables. Then, with taking into account the classical analog
of the unitary anisotropic rescaling transformation 
(\ref{canxpl}), one can show that classical solutions 
(\ref{traj}) and (\ref{Lis})
correspond to the
same  trajectories 
in four-dimensional phase space
projected  onto two different 
two-dimensional hyperplanes there which
correspond to coordinate variables $x_i$ and $x_i'$
of the ERIHO and AHO systems.

 \subsection{Relationship with harmonically extended Landau problem}\label{Landau}

Let us consider the Landau problem in symmetric gauge assuming  
that  the particle is subject to  the action of the additional quadratic potential term.
The Hamiltonian of the system is
\begin{eqnarray}
&\label{Lanext}
H_\pm=\frac{1}{2m}(\Pi^\pm_i)^2+\frac{1}{2}m \Lambda x_i^2\,,
\qquad \Pi^\pm_i=p_i- \frac{q}{c}A^\pm_i\,,
&
\end{eqnarray}
where $\Lambda\in \R$ is a constant of dimension $[t^{-2}]$ and $A^\pm_i$ is given in (\ref{Ag}).
Note that when $q=0$, Hamiltonian (\ref{Lanext}) 
with $\Lambda<0$ corresponds to the inverted isotropic harmonic oscillator.
If we choose the case of the positive upper index 
(that corresponds to $g=1$ in  (\ref{Ag})),  and expand this Hamiltonian, we obtain
\begin{eqnarray}
&
\label{LandauH}
H_+=\frac{1}{2m}(p_1^2+p_2^2)+\frac{m}{2}(\omega_B^{2}+ \Lambda)(x_1^2+x_2^2)+
\omega_B p_\varphi\,,\qquad
\omega_B=\frac{qB}{2mc}\,,
&
\end{eqnarray}
where now  we do not restrict the sign of $qB$.
Assuming that $\Lambda>-\omega_B^2$, i.e. that in the case of the inverted harmonic 
potential its coupling constant $|\Lambda | m $ is not  too   strong, we denote
 \be
 \label{ConditionOmega}
 \omega^2=\omega_B^{2}+\Lambda\,,\qquad \omega^{2}>0\,.
 \ee 
 With this restriction, the system is 
 confined: all its trajectories are bounded,  
 while the  
 quantum spectrum  is discrete.
 Introducing the parameter $g$ defined by relations
 $\sqrt{|\Lambda |}=\sqrt{|1-g^2|}\,\omega$,
 $\text{sign}\, (g)=\text{sign}\, (qB)$, 
 Hamiltonian (\ref{LandauH}) takes then the form 
 of the Hamiltonian of the
ERIHO system,
\begin{eqnarray}
&
H_+=H_{\text{osc}}+ g\omega p_\varphi =H_g\,.
&
\end{eqnarray}
Therefore, the case of Landau problem (in symmetric gauge)
subjected to the additional action of the isotropic harmonic potential trap
($\Lambda>0$)
is equivalent to the Euclidean case of the ERIHO system
with $g^2<1$, 
while the case  of not too strong 
inverted isotropic harmonic potential 
($0>\Lambda>-\omega_B^2$) 
is equivalent to the 
Minkowskian case of the ERIHO system
with $g^2>1$. 
Having this relationship of the ERIHO
systems $H_g$ 
with $g^2<1$ and $g^2>1$ with the Landau problem supplemented,
respectively,  
with the attractive and repulsive harmonic potential  terms,
one can understand the phenomenon of transmutation of symmetries
discussed in Sec. \ref{ClasPic}
in the light of different realizations  of conformal symmetries
considered in   \cite{Papan}.

In the critical case $\Lambda=-\omega_B^2 $, frequency $\omega$ turns into zero, 
and Hamiltonian (\ref{LandauH}) takes the form $H_c= \frac{1}{2m} p_ip_i + \omega_B p_\varphi$.
The system $H_c$ is not confined anymore,  its trajectories are
infinite,  and the corresponding quantum spectrum is continuous 
and not bounded from below. 
In supercritical case $\Lambda<-\omega^2_B$, $\omega^2$ is negative,  
and 
Hamiltonian (\ref{LandauH}) takes the form of  a   
two-dimensional inverted oscillator Hamiltonian
plus a Zeeman type term,
$H_{\text{sc}}=\frac{1}{2m}(p_1^2+p_2^2)-\frac{m}{2}|\omega^2|(x_1^2+x_2^2)+
\omega_B p_\varphi$.
All the peculiar properties of the critical 
case,  i.e.  infinite classical trajectories and continuous spectrum not bound
from below,  are inherited by $H_{\text{sc}}$. 

One can notice that  the critical case corresponds here to 
the dynamics of a free particle in a non-inertial, uniformly rotating 
reference frame that is described by the Lagrangian
\be\label{LOme}
L_\Omega=\frac{1}{2}m\left(\dot{\vec{r}}+\vec{\Omega}\times\vec{r}\right)^2.
\ee
Indeed, choosing $\vec{\Omega}=\frac{q}{2mc}\vec{B}$, and assuming that the uniform 
magnetic  field $\vec{B}$ is oriented perpendicular to the plane with coordinates
$x_1,\,x_2$, the dynamics in the direction orthogonal to this plane will be free.
Neglecting this free part of the dynamics,
Lagrangian (\ref{LOme}) can be reduced to 
\be\label{L+Ome}
L_\Omega^+=L_L+\frac{1}{2}m\omega_B^2(x_1^2+x_2^2)\,,
\qquad
L_L=\frac{1}{2}m(\dot{x}_1^2+\dot{x}_2^2)+\frac{q}{c}A^+_ix_i\,,
\ee
where $L_L$ is the Lagrangian of the Landau problem in
the symmetric gauge. Hamiltonian  (\ref{LandauH}) in the critical case 
of $\Lambda=-\omega_B^2$
corresponds exactly to the Lagrangian $L^+_\Omega$.
Then, our ERIHO system admits  yet another interpretation as a 
 plane isotropic  harmonic oscillator
 described by a potential $U=\frac{1}{2}k(x_1^2+x^2_2)$ in a 
non-inertial, uniformly rotating 
reference frame.
The cases $0<k<m\omega_B^2$, 
$k=m\omega_B^2$ and $k>m\omega_B^2$
correspond  to the phases   $|g|>1$, $|g|=1$ and  
$0<|g|<1$, respectively, with 
$\text{sing}\ g=\text{sing}\, \Omega_3$.
The inertial case  $\Omega=0$ with $k=m\omega^2$ corresponds, obviously,  to 
the isotropic Euclidean oscillator ($g=0$).
From this point of view, the Minkowskian phase, $g^2>1$,
corresponds to the case when the rotation frequency 
of the non-inertial reference frame 
dominates the oscillator frequency, $\Omega^2>k/m$,
while their equality, $\Omega^2=k/m$, corresponds to 
the Landau phases $g=\pm 1$.

\section{Discussion and Outlook}
\label{SecDisOut}

We studied  the ERIHO system that 
 represents an  isotropic Euclidean planar harmonic oscillator supplemented by
a kind of Zeeman-like term with  a dimensionless coupling  constant $g$.
The system was obtained  by generalizing  the conformal bridge 
transformation construction of refs.  \cite{InzPlyWipf1,InzPlyWipf2,InzPly7}
that  allows to relate harmonically confined models with associated asymptotically  
free systems.
To this aim, we applied a certain non-unitary  rotationally invariant
conformal intertwining operator  to the complex linear combination of the free
particle's    dilatation and rotation integrals.

We showed that the Hamiltonian of the obtained  ERIHO  system $H_g$
can be presented as a sum of the two  circular oscillatory modes
taken
with the relative weights $(1+g)$ and $(1-g)$.
As a consequence, the system 
reveals three different phases  depending on the
coupling constant value. 
In the case of $g^2<1$, the system represents 
the Euclidean phase of the ERIHO, that  turns into isotropic planar oscillator
at $g=0$.  The case $g^2>1$ corresponds to the Minkowskian phase of the ERIHO, 
which under  frequency rescaling 
$\omega\rightarrow \omega/|g|$ and taking the infinite 
limit $|g|\rightarrow\infty$
transforms into the isotropic, $\mathfrak{so}(2)$-invariant 
Minkowskian oscillator.
In the  cases $g=\pm 1$ 
the system reduces to the Landau problem in the symmetric gauge
with the opposite  orientation of the magnetic field.
The trajectories are closed for arbitrary choice of
the  initial data only for 
rational values of $g$.  
They 
have central symmetry 
except   the cases $g=\pm 1$
with arbitrary-centered circular orbits.
For $g=0$ and $|g|=\infty$, the  trajectories are, respectively,
 elliptic and circular.

The closed character of the trajectories at rational values of $g$ 
is reflected in the presence of the  hidden symmetry described by the 
pair of not depending explicitly on time, true integrals of motion that appear
in addition to the Hamiltonian $H_g$ and angular momentum $p_\varphi$.
The additional integrals are  quadratic in the circular oscillator variables in the    
cases $g=0$, $g^2=1$ and $g^2=\infty$.
Moreover,  in particular cases 
of $g=\pm 1$, they  
 are quadratic in the true linear integrals corresponding to the 
  translation symmetry generators which are the non-commuting 
  coordinates of the circumference' center.
In the indicated exceptional cases, the four true integrals 
together with other six  explicitly depending on time, dynamical integrals 
of the second order
generate the $\mathfrak{sp}(4,\R)$ Lie algebra, 
and the pairs of exceptional cases  $(g=0,\, |g|=\infty)$ and
$(g=+1,\,g=-1)$ are related by  a  transmutation  of symmetries in the following sense.
The $\mathfrak{sp}(4,\R)$ algebra of the isotropic Euclidean case 
($g=0$)
contains the $\mathfrak{u}(2)\cong\mathfrak{su}(2)\oplus\mathfrak{u}(1)$
subalgebra, generated by the four true integrals, and the 
$\mathfrak{gl}(2,\R)\cong\mathfrak{sl}(2,\R)\oplus\mathfrak{u}(1)$
algebra, in which $\mathfrak{sl}(2,\R)$ corresponds to the conformal symmetry 
while the center  $\mathfrak{u}(1)$ is generated by $p_\varphi$.
In the isotropic Minkowskian oscillator case ($|g|=\infty$), these symmetries
are interchanged:  the subalgebra $\mathfrak{gl}(2,\R)$
is generated by the true integrals, while $\mathfrak{u}(2)$
is associated with its conformal, dynamical symmetry.
In the cases $g=\pm1$  the
true and extended conformal symmetries generate
two copies of  the same subalgebra 
$\mathfrak{gl}(2,\R)$.
The generators of these subalgebars are interchanged under the change
$g=+1\leftrightarrow g=-1$.

In general case of rational values of $g$ the pair of additional true integrals
is of  higher order in circular oscillator variables. 
In the case of $g^2<1$ with 
$g=g^{s_1,s_2}_<=(s_2-s_1)/(s_2+s_1)$, where positive integer numbers 
$s_1$ and $s_2$ 
are chosen so that the fraction is irreducible, 
the pair of mutually complex conjugate additional 
integrals are of 
the  order $s_1+s_2$ in circular oscillator variables,
and together with $H_g$ and $p_\varphi$ they generate 
a non-linear deformation 
of the $\mathfrak{u}(2)$ algebra.
When  $g^2>1$ with 
$g=g^{s_1,s_2}_>=(s_2+s_1)/(s_2-s_1)=1/g^{s_1,s_2}_<$,
the corresponding additional true integrals have the same 
order $s_1+s_2$ in circular oscillator variables, but 
together  with $H_g$ and $p_\varphi$  they generate, instead,
 a non-linear deformation  of the $\mathfrak{gl}(2,\R)$.
 So,  in the case of rational $g$
  any system $H_g$ is maximally super-integrable 
  \cite{SuperIn1, SuperIn2}.
  Notice  that if to  change the notation $s_1\leftrightarrow s_2$ 
  in the case  $g^2>1$,  we  find  that the  corresponding generators of 
  hidden symmetries of the same orders $s_1+s_2$  in the phases with $g^2<1$ 
  and $g^2>1$ will mutually transmute under the inversion  $g\rightarrow -1/g$. 
  Then the statement on transmutation (duality) under the 
  inversion $g\rightarrow -1/g$ can also 
  be extended for the sets of 
  generators of the 
  true and extended conformal  symmetries 
  in the Landau problem 
  with the opposite orientation of the magnetic field
  ($g=+1$ and $g=-1$).

At the quantum level, the system with $g^2<1$ has a discrete positive spectrum
with finite degeneracy of energy levels.  The quantum analogs of 
the additional true integrals of motion control this degeneracy 
and allow to generate any state with a given energy value starting 
from any  eigenstate with the same value of energy.
In the case of rational $g$ with $g^2>1$, the picture is similar, but there spectrum is not bounded
from below and each energy level is infinitely degenerate.

On the other hand, we  showed
 how the usual AHO systems can be generated from the free particle by using 
a certain rotationally non-invariant, anisotropic  conformal bridge
transformation. In such systems, as  is well known, 
the trajectories are closed in the case of commensurable frequencies
and represent the Lissajous curves. The peculiarity 
of the AHO systems in comparison with the ERIHO  systems
is that in them the trajectories are the same in the cases of 
Euclidean and Minkowskian planar oscillators with the same values
of frequencies.
We showed that the  ERIHO and AHO systems 
with the corresponding parameter values can be related
by a unitary canonical transformation that
represents a composition of a certain $\mathfrak{su}(2)$ rotation
in an ``ambient" three-dimensional space and of 
anisotropic rescaling,
which is  an $\mathfrak{so}(1,1)\oplus \mathfrak{so}(1,1)$ Bogolyubov
transformation.

We also 
showed that the ERIHO systems are equivalent 
to the Landau problem in symmetric gauge subjected to the action
of the additional rotationally invariant harmonic potential term 
$\frac{1}{2}m \Lambda x_i^2$.
In this case the systems with positive  coupling constant $\Lambda>0$ 
correspond to the ERIHO systems  with $g^2<1$,
while the negative values $0> \Lambda> - (\frac{qB}{2mc})^2$, corresponding to 
the inverted oscillator in the subcritical  phase, 
yield the ERIHO systems with $g^2>1$. 
In the case of critical, $\Lambda= - (\frac{qB}{2mc})^2$
and supercritical values $\Lambda< - (\frac{qB}{2mc})^2$
of the coupling constant of the inverted harmonic potential term,
classical trajectories are infinite, and the quantum 
spectrum is continuous and not bounded from below.
These phases with critical and  super-critical values of the 
 inverted potential term coupling have no analogs in the
studied  by us  ERIHO systems. But  we notice here that
such phases appear   in the systems with 
exotic Newton-Hooke symmetries  and non-commutative geometry 
\cite{NH2,AGKP}. 

Finally, we  showed  that  
our ERIHO admits  yet another interpretation as a  
 plane isotropic  harmonic oscillator  in a 
 uniformly rotating  reference frame.
From this point of view, the phases $g^2<1$, $g^2=1$ and 
$g^2>1$ correspond, respectively, to the cases
$\Omega^2<k/m$, $\Omega^2=k/m$
and $\Omega^2>k/m$, where  ${\Omega}$ is the angular
frequency of the reference frame, and 
$k$ is the harmonic oscillator constant. 
Critical case of the harmonically extended Landau problem 
in this case corresponds to a free particle
in a  uniformly rotating  reference frame.

A phase transition in  rotating harmonically trapped 
Bose-Einstein condensates
is expected at  $\Omega^2=k/m$, see  \cite{Cooper,ALF,Goldman} and references therein.
It is interesting whether  the described peculiar properties 
related to hidden symmetries emerging 
at rational values of $g$ reveal themselves somehow 
in such systems. Taking into account  the 
unitary equivalence of the ERIHO and  AHO systems,
in this context  one also can  expect  that Bose-Einstein condensates in 
rotationally-non-invarint
harmonic  anisotropic traps \cite{BEC-ani,CBYu} should exhibit 
properties similar to those they 
have in rotating harmonic traps.

\vskip0.05cm
In conclusion, we note some problems that deserve attention for further research.
\vskip0.05cm

First of all, the considered CBTs of the rotationally invariant and
rotationally non-invariant nature  
 can be generalized  for dimensions higher than two.
Furthermore, the conformal potential term $ \gamma / (x_ix_i)$ 
can be included into the initial free particle Hamiltonian,
and analysis also can be extended for Calogero-type
systems  that would correspond to the case 
of anisotropic CBT.
Having in mind the constructions based  on exotic Galilean and 
Newton-Hooke
symmetries  \cite{AGKP,ZHAGK},
relevant to the physics of anyons \cite{NonCom2,NonComPLB},
it also would be interesting to apply our analysis 
to the case of non-commutative geometry.

In a recent paper \cite{InzPly7},  by  applying  the CBT to 
the study of  dynamics in a cosmic string background, 
we revealed  a kind of  quantum anomaly  there.
It would  be interesting to investigate  what happens with 
the quantum anomaly  in the presence of the Zeeman-like term.
 In fact, our interest in studying the ERIHO systems arose from an 
attempt to understand, in the light of symmetries, the 
Landau problem in a geometric background with topological defects 
such as a conical one \cite{KatVol,DeAMarques} 
 and a background of a rotating cosmic string \cite{Mazur,Muniz}.
 
\vskip0.2cm

{\bf {\large 
Acknowledgements}}

The work was partially supported by the FONDECYT Project 1190842 
and the DICYT Project 042131P\_POSTDOC.

\end{document}